\documentstyle[12pt]{article}  

\def\thefootnote{\fnsymbol{footnote}}
\def\ref#1{$^{#1)}$}

\newcommand{\be}{\begin{equation}}
\newcommand{\ee}{\end{equation}}
\newcommand{\bea}{\begin{eqnarray}}
\newcommand{\eea}{\end{eqnarray}}
\newcommand{\Lag}{{\cal L}}
\newcommand{\superint}{\int \diff^{4}\theta}
\newcommand{\lowest}{|_{\theta =\bar{\theta}=0}}
\newcommand{\diff}{\mbox{d}}
\newcommand{\Diff}{{\cal D}}

\newcommand{\DaDa}{{\cal D}^{\alpha}{\cal D}_{\alpha}}
\newcommand{\DbDb}{{\cal D}_{\dot{\alpha}}{\cal D}^{\dot{\alpha}}}

\newcommand{\hs}{\hspace{0.2cm}}
\newcommand{\dilaton}{{\ell}}
\newcommand{\ldgl}{\dilaton g_{_{\dilaton}}}
\newcommand{\zdgz}{z g_{_{z}}}
\newcommand{\xdgx}{x g_{_{x}}}
\newcommand{\ldfl}{\dilaton f_{_{\dilaton}}}
\newcommand{\zdfz}{z f_{_{z}}}
\newcommand{\xdfx}{x f_{_{x}}}

\newcommand{\ldhl}{\dilaton h_{_{\dilaton}}}
\newcommand{\zdhz}{z h_{_{z}}}
\newcommand{\xdhx}{x h_{_{x}}}
\newcommand{\axion}{a}
\newcommand{\ep}{\varepsilon}

\begin{document}
\begin{titlepage}
\begin{center}
            \hfill    LBNL-39441 \\
            \hfill    UCB-PTH-96/42 \\
            \hfill    hep-th/9610089 \\[0.0in]
{\large \bf Dilaton Stabilization and Supersymmetry Breaking\\
        by Dynamical Gaugino Condensation in the Linear\\
        Multiplet Formalism of String Effective Theory}
\footnote{This work was supported in part by the Director, Office of 
Energy Research, Office of High Energy and Nuclear Physics, Division 
of High Energy Physics of the U.S. Department of Energy under 
Contract DE-AC03-76SF00098 and in part by the National Science 
Foundation under grant PHY-95-14797.}\\[.07in]

                 {\bf Yi-Yen Wu} \\[.05in]

{\em  Theoretical Physics Group \\
      Ernest Orlando Lawrence Berkeley National Laboratory \\
      University of California, Berkeley, California 94720 \\
      and \\
      Department of Physics \\
      University of California, Berkeley, California 94720}\\[.07in]
\end{center}

\begin{abstract}
\vspace{-0.04in}
We study dynamical gaugino condensation in superstring effective theories 
using the linear multiplet representation for the dilaton superfield. An
interesting necessary condition for the dilaton to be stabilized, which was
first derived in generic models of static gaugino condensation, is shown to 
hold for generic models of dynamical gaugino condensation. We also point out 
that it is stringy non-perturbative effects that stabilize the dilaton and 
allow dynamical supersymmetry breaking via the field-theoretical 
non-perturbative effect of gaugino condensation. As a typical example, a toy 
S-dual model of a dynamical E$_{8}$ condensate is constructed and the dilaton 
is explicitly shown to be stabilized with broken supersymmetry and (fine-tuned) 
vanishing cosmological constant.
\end{abstract}
\end{titlepage}
\renewcommand{\thepage}{\roman{page}}
\setcounter{page}{2}
\mbox{ }

\vskip 1in

\begin{center}
{\bf Disclaimer}
\end{center}

\vskip .2in

\begin{scriptsize}
\begin{quotation}
This document was prepared as an account of work sponsored by the United
States Government.  Neither the United States Government nor any agency
thereof, nor The Regents of the University of California, nor any of their
employees, makes any warranty, express or implied, or assumes any legal
liability or responsibility for the accuracy, completeness, or usefulness
of any information, apparatus, product, or process disclosed, or represents
that its use would not infringe privately owned rights.  Reference herein
to any specific commercial products process, or service by its trade name,
trademark, manufacturer, or otherwise, does not necessarily constitute or
imply its endorsement, recommendation, or favoring by the United States
Government or any agency thereof, or The Regents of the University of
California.  The views and opinions of authors expressed herein do not
necessarily state or reflect those of the United States Government or any
agency thereof of The Regents of the University of California and shall
not be used for advertising or product endorsement purposes.
\end{quotation}
\end{scriptsize}

\vskip 2in

\begin{center}
\begin{small}
{\it Lawrence Berkeley Laboratory is an equal opportunity employer.}
\end{small}
\end{center}

\newpage
\renewcommand{\thepage}{\arabic{page}}
\renewcommand{\theequation}{\arabic{section}.\arabic{equation}}
\def\thefootnote{\arabic{footnote}}
\setcounter{page}{1}
\section{Introduction}
\hspace{0.8cm}\setcounter{equation}{0}
Constructing a realistic scheme of supersymmetry breaking is one of the big
challenges to supersymmetry phenomenology. However, in the context of
superstring phenomenology, there are actually more challenges. As is well 
known, a very powerful feature of superstring phenomenology is that all 
the {\em parameters} of the model are in principle dynamically determined by 
the {\em vev's} of certain fields. One of these important fields is the 
string dilaton whose {\em vev} determines the gauge coupling constants. On the
other hand, how the dilaton is stabilized is outside the reach of perturbation
theory since the dilaton's potential remains flat to all order in perturbation
theory according to the non-renormalization theorem. Therefore, understanding 
how the dilaton is stabilized (i.e., how the gauge coupling constants are 
determined) is of no less significance than understanding how supersymmetry is 
broken.

   Gaugino condensation has been playing a unique role in these issues: At low
energy, the strong dilaton-Yang-Mills interaction leads to gaugino condensation 
which not only breaks supersymmetry spontaneously but also generates a
non-perturbative dilaton potential which may eventually stabilize the dilaton. 
In the scheme of gaugino condensation the stabilization of the dilaton and the
breaking of supersymmetry are therefore unified in the sense that they are 
two aspects of a single non-perturbative phenomenon. Furthermore, gaugino
condensation has its own important phenomenological motivations: gaugino
condensation occurs in the hidden sector of a generic string model 
\cite{Nilles82,Dine85}; it can break supersymmetry at a sufficiently small 
scale and induce viable soft supersymmetry breaking effects in the observable 
sector through gravity and/or an anomalous U(1) gauge interaction 
\cite{messenger}.

   Unfortunately, this beautiful scheme of gaugino condensation has been long
plagued by the infamous dilaton runaway problem \cite{Dine85,Banks94}. That 
is, (assuming that the tree-level K\"ahler potential of the dilaton is a good
approximation) one generally finds that the supersymmetric vacuum with vanishing
gauge coupling constant and no gaugino condensation is the only stable minimum
in the weak-coupling regime. (The recent observation of string dualities further
implies that the strong-coupling regime is plagued by a similar runaway problem 
\cite{stringduality}.) Only a few solutions to the dilaton runaway problem have
been proposed. Assuming the scenario of two or more gaugino condensates, the 
racetrack model stabilizes the dilaton and breaks supersymmetry with a more
complicated dilaton superpotential generated by multiple gaugino condensation
\cite{racetrack}. However, stabilization of the dilaton in the racetrack 
model requires a delicate cancellation between the contributions from different
gaugino condensates, which is not very natural. Furthermore, it has a large and
negative cosmological constant when supersymmetry is broken. The other
solutions generically require the assistance of an additional source of 
supersymmetry breaking (e.g., a constant term in the superpotential) 
\cite{Dine85,constant}. It is therefore fair to say that there is no 
satisfactory solution so far.

Recently, several new developments and insights of superstring phenomenology are
now known to play important roles in the above issues, and it is the purpose of
this paper to show how these new ingredients can eventually lead to a promising
solution. One of these new ingredients is the linear multiplet formalism of
superstring effective theories \cite{Adamietz93,Derendinger94}: the dilaton
superfield can be described either by a chiral superfield $S$ or by a linear 
multiplet $L$ \cite{Linear}, which is known as the chiral-linear duality. Since
the precise field content of the linear multiplet $L$ appears in the massless 
string spectrum and $\langle L\rangle$ plays the role of string loop expansion 
parameter, stringy information is more naturally encoded in the linear 
multiplet formalism rather than in the chiral formalism. As will be pointed 
out later, stringy effects are believed to be important in the 
stabilization of the dilaton and supersymmetry breaking by gaugino condensation;
therefore, it is more appropriate to study these issues in the linear 
multiplet formalism.

The other new ingredient concerns the effective description of gaugino 
condensation. In the known models of gaugino condensation using the chiral 
superfield representation for the dilaton, the gaugino condensate has always 
been described by an {\em unconstrained} chiral superfield $U$ which 
corresponds to the bound state of $\,{\cal W}^{\alpha}{\cal W}_{\alpha}\,$ in 
the underlying theory. It was pointed out recently that $U$ should be a 
{\em constrained} chiral superfield \cite{Binetruy95,sduality,Burgess95,Pillon} 
due to the constrained superspace geometry of the underlying Yang-Mills theory:
\bea
U\,&=&\,-(\DbDb-8R)V, \nonumber \\ 
\bar{U}\,&=&\,-(\DaDa-8R^{\dagger})V,
\eea
where $V$ is an unconstrained vector superfield. Furthermore, in the linear 
multiplet formalism the linear multiplet $L$ and the constrained $U$, $\bar{U}$
nicely merge into an unconstrained vector superfield $V$, and therefore the 
effective Lagrangian can elegantly be described by $V$ alone.

The third new ingredient is the stringy non-perturbative effect conjectured by
\mbox{S.H. Shenker} \cite{Shenker90}. It is further argued in \cite{Banks94} 
that the K\"ahler potential can in principle receive significant stringy
non-perturbative corrections although the superpotential cannot generically. 
Significant stringy non-perturbative corrections to the K\"ahler potential imply
that the usual dilaton runaway picture is valid only in the weak-coupling
regime; as pointed out in \cite{Banks94}, these corrections may naturally
stabilize the dilaton.\footnote{Choosing a specific form for possible 
non-perturbative corrections to the K\"ahler potential, \cite{Casas96} has
discussed the possibility of stabilizing the dilaton in a model of gaugino 
condensation using chiral superfield representation for the dilaton. 
However, the issue of modular anomaly cancellation was not taken into account.} 
However, it may seem futile discussing whether the dilaton can be stabilized 
or not because we do not know what these non-perturbative corrections to the 
K\"ahler potential really are. On the other hand, in the spirit of 
effective Lagrangian approach it is always legitimate to ask the following 
interesting questions without having to specify the non-perturbative 
corrections to the K\"ahler potential: What is the generic condition for the 
dilaton to be stabilized? Is supersymmetry broken if the dilaton is stabilized?
By studying generic models of static gaugino condensation with the above 
three new ingredients included, an interesting necessary condition for the 
dilaton to be stabilized has been derived in \cite{dilaton}. It is also shown 
that supersymmetry is broken as long as the dilaton is stabilized. Furthermore, 
explicit models with stabilized dilaton, broken supersymmetry and (fine-tuned) 
vanishing vacuum energy can be constructed.

As pointed out in \cite{sduality}, the kinetic terms for gaugino 
condensate naturally arise both from field-theoretical loop corrections and from
classical string corrections \cite{Ant}. Therefore, in this paper 
we would like to address the above questions in the context of dynamical
gaugino condensation. In \mbox{Sect. 2}, the field component Lagrangian for the 
generic model of dynamical gaugino condensation is constructed, and its vacuum
structure is analyzed. In \mbox{Sect. 3}, we review the study of static gaugino 
condensation \cite{dilaton} which is essential to the study of dynamical 
gaugino condensation later. The role of stringy non-perturbative effects in 
stabilizing the dilaton is also discussed. In \mbox{Sect. 4}, the S-dual models 
of dynamical gaugino condensation are studied. We discuss how the model of 
static gaugino condensation is related to the model of dynamical gaugino 
condensation and its implications. It is shown that the necessary condition of 
dilaton stabilization derived from static gaugino condensation also holds for 
generic models of dynamical gaugino condensation. 
\section{Generic Model of Dynamical Gaugino Condensation}
\setcounter{equation}{0} 
\hspace{0.8cm}
It will be shown in this section how to construct the component field 
Lagrangian for the generic model of dynamical gaugino condensation using the 
K\"ahler superspace formalism of supergravity \cite{Binetruy90,Binetruy91}.
We consider here orbifold models with gauge groups 
E$_{8}\otimes$E$_{6}\otimes$U(1)$^{2}$, three untwisted (1,1) moduli $T^{I}$ 
($I\,=\,1,\;2,\;3$) \cite{Gaillard92,KL,Dixon90}, and universal modular anomaly
cancellation \cite{ant} (e.g., the Z$_{3}$ and Z$_{7}$ orbifolds). The confined
E$_{8}$ hidden sector is described by the following generic model of a single 
dynamical gaugino condensate $U$ with K\"ahler potential $K$:
\bea
K\,&=&\,\ln V\,+\,g(V,\bar{U}U)\,+\,G, \nonumber \\
\Lag_{eff}\,&=&\,
\superint\,E\,\left\{\,\left(\,-2\,+\,f(V,\bar{U}U)\,\right)\,+\,bVG\,\right\} 
\,+\,\left\{\,\superint\,\frac{E}{R}\,e^{K/2}W_{VY}\,+\,\mbox{h.c.}\,\right\},
\nonumber \\
G\,&=&\,-\sum_{I}\ln(T^{I}+\bar{T}^{I}),
\eea
where $\;U\,=\,-(\DbDb-8R)V\,$, $\,\bar{U}\,=\,-(\DaDa-8R^{\dagger})V.\;$ We 
also write $\;\ln V\,+\,g(V,\bar{U}U)\,\equiv\,k(V,\bar{U}U).\;$ The term
$\,\left(\,-2\,+\,f(V,\bar{U}U)\,\right)\,$ of $\Lag_{eff}$ is the superspace 
integral which yields the kinetic actions for the linear multiplet, 
supergravity, matter, and gaugino condensate. The term $\,bVG\,$ is the
Green-Schwarz counterterm \cite{Gaillard92} which cancels the full modular 
anomaly here. $b\,=\,C/8\pi^{2}\,=\,2b_{0}/3$, and $C\,=\,30$ is the Casimir 
operator in the adjoint representation of $\mbox{E}_{8}$. $b_{0}$ is the 
$\mbox{E}_{8}$ one-loop $\beta$-function coefficient. $W_{VY}$ is the quantum
superpotential whose form is dictated by the underlying anomaly structure
\cite{Veneziano82}:
\be
\superint\,\frac{E}{R}\,e^{K/2}W_{VY}\,=\,
\superint\,\frac{E}{R}\,\frac{1}{8}bU\ln(e^{-K/2}U/\mu^{3}),
\ee
where $\mu$ is a constant left undetermined by anomaly matching. 
$g(V,\bar{U}U)$ and $f(V,\bar{U}U)$ represent the quantum corrections to the 
tree-level K\"{a}hler potential. As illustrated in \mbox{Sect. 1}, 
$\,g(V,\bar{U}U)\,$ and $\,f(V,\bar{U}U)\,$ are taken to be arbitrary but 
bounded here. The dynamical model (2.1) is the straightforward generalization 
of the static model in \cite{dilaton} by including the $\bar{U}U$ dependence 
in the K\"ahler potential. We can also rewrite (2.1) as a single D term:
\bea
K\,&=&\,\ln V\,+\,g(V,\bar{U}U)\,+\,G, \nonumber \\
\Lag_{eff}\,&=&\,
\superint\,E\,\left\{\,\left(\,-2\,+\,f(V,\bar{U}U)\,\right)\,+\, 
bVG\,+\,bV\ln(e^{-K}\bar{U}U/\mu^{6})\,\right\}.\hspace{1.5cm}
\eea

Throughout this paper only the bosonic and gravitino parts of the component
field Lagrangian are presented. In the following, we enumerate the definitions 
of the bosonic component fields:
\bea
\dilaton\,&=&\,V\lowest,\nonumber\\
\sigma^{m}_{\alpha\dot{\alpha}}B_{m}\,&=&\,
\frac{1}{2}\left[\,\Diff_{\alpha},\Diff_{\dot{\alpha}}\,\right]V\lowest\,+\,
\frac{2}{3}\dilaton\sigma^{a}_{\alpha\dot{\alpha}} b_{a},\nonumber\\
u\,&=&\,U\lowest\,=\,-(\bar{\Diff}^{2}-8R)V\lowest,\nonumber\\
\bar{u}\,&=&\,\bar{U}\lowest\,=\,-(\Diff^{2}-8R^{\dagger})V\lowest,
\nonumber \\
-4F_{U}\,&=&\,\Diff^{2}U\lowest, \;\;\; 
-4\bar{F}_{\bar{U}}\,=\,\bar{\Diff}^{2}\bar{U}\lowest,
\nonumber \\
D\,&=&\,\frac{1}{8}\Diff^{\beta}(\bar{\Diff}^{2}-8R)
      \Diff_{\beta}V\lowest\nonumber\\
   &=&\,\frac{1}{8}\Diff_{\dot{\beta}}(\Diff^{2}-8R^{\dagger})
      \Diff^{\dot{\beta}}V\lowest,
\nonumber \\
t^{I}\,&=&\,T^{I}\lowest,\;\;\;
-4F_{T}^{I}\,=\,\Diff^{2}T^{I}\lowest,\nonumber\\
\bar{t}^{I}\,&=&\,\bar{T}^{I}\lowest,\;\;\;
-4\bar{F}_{\bar{T}}^{I}\,=\,\bar{\Diff}^{2}\bar{T}^{I}\lowest,
\eea
where $\,b_{a}\,=\,-3G_{a}\lowest\,$, $\,M\,=\,-6R\lowest\,$, 
$\,\bar{M}\,=\,-6R^{\dagger}\lowest\,$ are the auxiliary components of the 
supergravity multiplet. $(F_{U}-\bar{F}_{\bar{U}})$ can be expressed as follows:
\be
(F_{U}-\bar{F}_{\bar{U}})\,=\,4i\nabla^{m}\!B_{m}
\,+\,u\bar{M}\,-\,\bar{u}M,
\ee
and $(F_{U}+\bar{F}_{\bar{U}})$ contains the auxiliary field $D$. We also 
write $\,Z\,\equiv\,\bar{U}U\,$, and its bosonic component 
$\,z\,\equiv\,Z\lowest\,=\,\bar{u}u\,$.

The construction of component field Lagrangian using chiral density 
multiplet method~\cite{Binetruy90} has been detailed in \cite{dilaton}, and
therefore only the key steps are presented here. The chiral density multiplet 
${\bf r}$ and its hermitian conjugate ${\bf \bar{r}}$ for the generic 
model (2.1) are:
\bea
{\bf r}\,&=&\,-\,\frac{1}{8}(\bar{\Diff}^{2}-8R)
\left\{\,\left(\,-2\,+\,f(V,\bar{U}U)\,\right)
\,+\,bVG\,+\,bV\ln(e^{-K}\bar{U}U/\mu^{6})\,\right\}, \nonumber\\
{\bf \bar{r}}\,&=&\,-\,\frac{1}{8}(\Diff^{2}-8R^{\dagger})
\left\{\,\left(\,-2\,+\,f(V,\bar{U}U)\,\right)
\,+\,bVG\,+\,bV\ln(e^{-K}\bar{U}U/\mu^{6})\,\right\},
\hspace{1.5cm}
\eea
and the component field Lagrangian $\Lag_{eff}$ is:
\bea
\frac{1}{e}\Lag_{eff}\,&=&\,-\,\frac{1}{4}\Diff^{2}{\bf r}
\lowest\,+\,\frac{i}{2}(\bar{\psi}_{m}\bar{\sigma}^{m})^{\alpha}
\Diff_{\alpha}{\bf r}\lowest\nonumber\\
& &\,-\,(\bar{\psi}_{m}\bar{\sigma}^{mn}\bar{\psi}_{n}+\bar{M})
{\bf r}\lowest \,\,+\,\, \mbox{h.c.}
\eea
We choose to write out explicitly the vectorial part of the K\"{a}hler
connection $\,A_{m}\,$ and keep only the Lorentz connection in the definition 
of the covariant derivatives when we present component expressions. 
The $\: A_{m}\lowest\:$ for the generic model (2.1) is:
\bea
A_{m}\lowest\,&=&\,-\,\frac{i}{4\dilaton}\!\cdot\!
\frac{(1+\ldgl)}{(1-\zdgz)}B_{m}\,+\,\frac{i}{6}
\left[\,\frac{(1+\ldgl)}{(1-\zdgz)}\,-\,3\,\right]e_{m}^{\hs a}b_{a}
\nonumber\\
& &\,+\,\frac{1}{4(1-\zdgz)}\sum_{I}\frac{1}{(t^{I}+\bar{t}^{I})}
(\nabla_{\!m}\bar{t}^{I}\,-\,\nabla_{\!m}t^{I})
\nonumber\\
& &\,-\,\frac{\zdgz}{4(1-\zdgz)}\nabla_{\!m}\ln\!\left(\frac{\bar{u}}{u}\right).
\eea
The following are the simplified notations for partial derivatives of $g$:
\be
g_{_{\dilaton}}\,\equiv\,\frac{\partial g(\dilaton,z)}{\partial \dilaton},
\;\;\;g_{_{z}}\,\equiv\,\frac{\partial g(\dilaton,z)}{\partial z},
\ee
and similarly for other functions.

In the computation of (2.7), we need to decompose the lowest 
components of the following six superfields:
$X_{\alpha}$, $\bar{X}^{\dot{\alpha}}$, $\Diff_{\alpha}R$,
$\Diff^{\dot{\alpha}}\!R^{\dagger}$, 
$(\Diff^{\alpha}\!X_{\alpha}+\Diff_{\dot{\alpha}}\bar{X}^{\dot{\alpha}})$
and $(\Diff^{2}\!R+\bar{\Diff}^{2}\!R^{\dagger})$ into component 
fields, where
\bea
X_{\alpha}\,&=&\,-\,\frac{1}{8}(\DbDb-8R)
\Diff_{\alpha}K,
\nonumber\\
\bar{X}^{\dot{\alpha}}\,&=&\,-\,\frac{1}{8}(\DaDa-8R^{\dagger})
\Diff^{\dot{\alpha}}\!K,
\nonumber\\
(\Diff^{\alpha}\!X_{\alpha}+\Diff_{\dot{\alpha}}\bar{X}^{\dot{\alpha}})\,
&=&\,-\,\frac{1}{8}\Diff^{2}\!\bar{\Diff}^{2}\!K
\,-\,\frac{1}{8}\bar{\Diff}^{2}\!\Diff^{2}\!K
\,-\,\Diff^{\alpha\dot{\alpha}}\!\Diff_{\alpha\dot{\alpha}}K
\nonumber\\
& &\,-\,G^{\alpha\dot{\alpha}}
\left[\,\Diff_{\alpha},\Diff_{\dot{\alpha}}\,\right]K
\,+\,2R^{\dagger}\bar{\Diff}^{2}\!K\,+\,2R\Diff^{2}\!K
\nonumber\\
& &\,-\,(\,\Diff^{\alpha}\!G_{\alpha\dot{\alpha}}
\,-\,2\Diff_{\dot{\alpha}}R^{\dagger}\,)\Diff^{\dot{\alpha}}\!K
\nonumber\\
& &\,+\,(\,\Diff^{\dot{\alpha}}\!G_{\alpha\dot{\alpha}}
\,+\,2\Diff_{\alpha}R\,)\Diff^{\alpha}\!K.
\eea
This is done by solving the following six algebraic equations: 
\bea
\left(1+V\frac{\partial g}{\partial V}\right)\Diff_{\alpha}R\,+\,
\left(1-Z\frac{\partial g}{\partial Z}\right)X_{\alpha}\,&=&\,\Xi_{\alpha},\\
3\Diff_{\alpha}R\,+\,X_{\alpha}\,&=&\,
-2(\sigma^{cb}\epsilon)_{\alpha\varphi}T_{cb}^
{\hs\hspace{0.04cm}\varphi}.
\eea
\bea
\left(1+V\frac{\partial g}{\partial V}\right)\Diff^{\dot{\alpha}}\!R^{\dagger}
\,+\,\left(1-Z\frac{\partial g}{\partial Z}\right)\bar{X}^{\dot{\alpha}}\,&=&\,
\bar{\Xi}^{\dot{\alpha}},\\
3\Diff^{\dot{\alpha}}\!R^{\dagger}\,+\,\bar{X}^{\dot{\alpha}}\,&=&\,
-2(\bar{\sigma}^{cb}\epsilon)^{
\dot{\alpha}\dot{\varphi}}T_{cb\dot{\varphi}}.
\eea
\bea
\left(1+V\frac{\partial g}{\partial V}\right)(\Diff^{2}\!R+\bar{\Diff}^{2}\!
R^{\dagger})\,+\,\left(1-Z\frac{\partial g}{\partial Z}\right)
(\Diff^{\alpha}\!X_{\alpha}+
\Diff_{\dot{\alpha}}\bar{X}^{\dot{\alpha}})\,&=&\,\Delta,\\
3(\Diff^{2}\!R+\bar{\Diff}^{2}\!R^{\dagger})\,+\,(\Diff^{\alpha}
\!X_{\alpha}+\Diff_{\dot{\alpha}}\bar{X}^{\dot{\alpha}})\,&=&\,
-2R_{ba}^{\hs ba}\,+\,12G^{a}G_{a} \hspace{0cm}\nonumber\\
\,& &\,+\,96RR^{\dagger}. \hspace{0cm}
\eea
The identities (2.12), (2.14) and (2.16) arise from the structure of K\"{a}hler
superspace. The identities (2.11), (2.13) and (2.15) arise from the definitions
of $X_{\alpha}$, $\bar{X}^{\dot{\alpha}}$, and
$(\Diff^{\alpha}\!X_{\alpha}+\Diff_{\dot{\alpha}}\bar{X}^{\dot{\alpha}})$. 
The computation of (2.10) defines the contents of $\Xi_{\alpha}$,
$\bar{\Xi}^{\dot{\alpha}}$ and $\Delta$. Eqs.(2.8-16) describe the key steps
in the computations of (2.7). In the following subsections, several important 
issues of this construction will be discussed.
\subsection{Canonical Einstein Term}
\hspace{0.8cm}
In order to have the correctly normalized Einstein term in $\Lag_{eff}$, an
appropriate constraint should be imposed on the generic model (2.1). Therefore,
it is shown below how to compute the Einstein term for (2.1). According to 
(2.7), the following are those terms in $\Lag_{eff}$ that will contribute to 
the Einstein term:
\bea
\frac{1}{e}\Lag_{eff}\;&\ni&\;\frac{1}{4}
\left[\,2-f+\ldfl-b\dilaton(1+\ldgl)\,\right]
(\Diff^{2}\!R+\bar{\Diff}^{2}\!R^{\dagger})\lowest
\nonumber\\
\;& &\;+\,\frac{1}{32}
\left[\,\zdfz+b\dilaton(1-\zdgz)\,\right]
\left(\,\frac{1}{\bar{u}}\Diff^{2}\!\bar{\Diff}^{2}\!\bar{U}
\,+\,\frac{1}{u}\bar{\Diff}^{2}\!\Diff^{2}\!U\,\right)\lowest.\hspace{1.5cm}
\eea
Note that the terms $\,\Diff^{2}\!\bar{\Diff}^{2}\!\bar{U}\,$ and 
$\,\bar{\Diff}^{2}\!\Diff^{2}\!U\,$ are related to 
$\,\Diff^{\alpha}\!X_{\alpha}\,$ and  
$\,\Diff_{\dot{\alpha}}\bar{X}^{\dot{\alpha}}\,$ through the following 
identities:
\bea
\Diff^{2}\!\bar{\Diff}^{2}\!\bar{U}\,&=&\,
16\Diff^{a}\!\Diff_{a}\bar{U}\,+\,64iG^{a}\Diff_{a}\bar{U}
\,-\,48\bar{U}G^{a}\!G_{a}\,+\,48i\bar{U}\Diff^{a}\!G_{a} \nonumber\\
\,& &\,-\,8\bar{U}\Diff^{\alpha}\!X_{\alpha}
\,+\,16R^{\dagger}\bar{\Diff}^{2}\!\bar{U}
\,+\,8(\Diff^{\alpha}\!G_{\alpha\dot{\alpha}})(\Diff^{\dot{\alpha}}\!\bar{U}).
\nonumber\\
\bar{\Diff}^{2}\!\Diff^{2}\!U\,&=&\,
16\Diff^{a}\!\Diff_{a}U\,-\,64iG^{a}\Diff_{a}U
\,-\,48UG^{a}\!G_{a}\,-\,48iU\Diff^{a}\!G_{a} \nonumber\\
\,& &\,-\,8U\Diff_{\dot{\alpha}}\bar{X}^{\dot{\alpha}}
\,+\,16R\Diff^{2}\!U
\,-\,8(\Diff^{\dot{\alpha}}\!G_{\alpha\dot{\alpha}})(\Diff^{\alpha}\!U).
\eea
The contributions of $\,(\Diff^{2}\!R+\bar{\Diff}^{2}\!R^{\dagger})\lowest\,$
and $\,(\Diff^{\alpha}\!X_{\alpha}+\Diff_{\dot{\alpha}}\bar{X}^{\dot{\alpha}})
\lowest\,$ to the Einstein term are obtained by solving (2.15-16):
\bea
(\Diff^{2}\!R+\bar{\Diff}^{2}\!R^{\dagger})\lowest\,&\ni&\,
-\,\frac{2(1-\zdgz)}{(2-\ldgl-3\zdgz)}R_{ba}^{\hs ba}\lowest.
\nonumber\\
(\Diff^{\alpha}\!X_{\alpha}+\Diff_{\dot{\alpha}}\bar{X}^{\dot{\alpha}})
\lowest\,&\ni&\,
+\,\frac{2(1+\ldgl)}{(2-\ldgl-3\zdgz)}R_{ba}^{\hs ba}\lowest.
\eea

By combining (2.17-19), it is straightforward to show that the Einstein term in 
$\Lag_{eff}$ is correctly normalized if and only if the following constraint 
is imposed:
\be
(\,1\,+\,\zdfz\,)(\,1\,+\,\ldgl\,)\,=\,
(\,1\,-\,\zdgz\,)(\,1\,-\,\ldfl\,+\,f\,),
\ee
which is a first-order partial differential equation. From now on, the study of 
the generic model (2.1) always assumes the constraint (2.20). (2.20) will be 
useful in simplifying the expression of $\Lag_{eff}$, and it turns out to 
be convenient to define $h$ as follows:
\bea
h\,&\equiv&\,\frac{(\,1\,+\,\zdfz\,)}{(\,1\,-\,\zdgz\,)},\nonumber\\
&=&\,\frac{(\,1\,-\,\ldfl\,+\,f\,)}{(\,1\,+\,\ldgl\,)}.
\eea
Furthermore, the partial derivatives of $h$ satisfy the following consistency
condition:
\be
(\,h\,-\,\ldhl\,)(\,\zdgz\,-\,1\,)\,+\,\zdhz(\,1\,+\,\ldgl\,)\,+\,1\,=\,0.
\ee
Eqs.(2.21-22) will also be very useful in simplifying the expression of 
$\Lag_{eff}$. Notice that $h=1$ for generic models of static gaugino 
condensation, and (2.20) is reduced to an ordinary differential equation 
\cite{dilaton}. We will show in \mbox{Sect. 4.2} how to construct physically
interesting solutions for the partial differential equation (2.20). 
\subsection{Component Field Lagrangian with Auxiliary Fields}
\hspace{0.8cm}
Once the issue of canonical Einstein term is settled, it is straightforward to
compute $\Lag_{eff}$ according to (2.8-16). The rest of it is standard and will
not be detailed here. In the following, we present the component field 
expression of $\Lag_{eff}$ as the sum of the bosonic Lagrangian $\Lag_{B}$ and 
the gravitino Lagrangian $\Lag_{\tilde{G}}$.
\be
\Lag_{eff}\,=\,\Lag_{B}\,+\,\Lag_{\tilde{G}}.
\ee

\bea
\frac{1}{e}\Lag_{B}\,&=&\,-\,\frac{1}{2}{\cal R}
\,-\,\frac{1}{4\dilaton^{2}}(h-\ldhl)(1+\ldgl)
\nabla^{m}\!\dilaton\,\nabla_{\!m}\!\dilaton
\nonumber\\
& &\,+\,\frac{1}{2\dilaton}\zdhz(1+\ldgl)\,
\nabla^{m}\!\ln(\bar{u}u)\,\nabla_{\!m}\!\dilaton
\nonumber\\
& &\,+\,\frac{u}{4\bar{u}}h_{_{z}}\!\cdot\!\zdgz
\frac{(2-\zdgz)}{(1-\zdgz)}
\nabla^{m}\bar{u}\,\nabla_{\!m}\bar{u}
\nonumber\\
& &\,-\,\frac{1}{2}h_{_{z}}
\left[\frac{(2-\zdgz)}{(1-\zdgz)}\,-\,\zdgz\right]
\nabla^{m}\bar{u}\,\nabla_{\!m}u
\nonumber\\
& &\,+\,\frac{\bar{u}}{4u}h_{_{z}}\!\cdot\!\zdgz
\frac{(2-\zdgz)}{(1-\zdgz)}
\nabla^{m}u\,\nabla_{\!m}u
\nonumber\\
& &\,-\,\frac{\zdhz}{2(1-\zdgz)}\sum_{I}\frac{1}{(t^{I}+\bar{t}^{I})}
(\,\nabla^{m}\bar{t}^{I}\,-\,\nabla^{m}t^{I}\,)\,
\nabla_{\!m}\!\ln\!\left(\frac{\bar{u}}{u}\right)
\nonumber\\
& &\,+\,\frac{\zdhz}{4(1-\zdgz)}
\sum_{I,J}\frac{1}{(t^{I}+\bar{t}^{I})(t^{J}+\bar{t}^{J})}
\nabla^{m}\bar{t}^{I}\,\nabla_{\!m}\bar{t}^{J}
\nonumber\\
& &\,-\,\frac{1}{2}\sum_{I,J}\left[2(h+b\dilaton)\delta_{IJ}\,+\,
\frac{\zdhz}{(1-\zdgz)}\right]
\frac{\nabla^{m}\bar{t}^{I}\,\nabla_{\!m}t^{J}}
{(t^{I}+\bar{t}^{I})(t^{J}+\bar{t}^{J})}
\nonumber\\
& &\,+\,\frac{\zdhz}{4(1-\zdgz)}
\sum_{I,J}\frac{1}{(t^{I}+\bar{t}^{I})(t^{J}+\bar{t}^{J})}
\nabla^{m}t^{I}\,\nabla_{\!m}t^{J}
\nonumber\\
& &\,+\,\frac{(2-\ldgl-3\zdgz)}{9(1-\zdgz)}\,b^{a}b_{a}
\nonumber\\
& &\,+\,\frac{(1+\ldgl)}{4\dilaton^{2}(1-\zdgz)}B^{m}\!B_{m}
\nonumber\\
& &\,+\,\frac{i}{2\dilaton}\left[h\,+\,b\dilaton\,-\,
\frac{1}{(1-\zdgz)}\right]\,B^{m}\,
\nabla_{\!m}\!\ln\!\left(\frac{\bar{u}}{u}\right)
\nonumber\\
& &\,-\,\frac{i}{2\dilaton}\left[h\,+\,b\dilaton\,-\,
\frac{1}{(1-\zdgz)}\right]\sum_{I}
\frac{(\,\nabla^{m}\bar{t}^{I}\,-\,\nabla^{m}t^{I}\,)}
{(t^{I}+\bar{t}^{I})}B_{m}
\nonumber\\
& &\,+\,4h_{_{z}}(1-\zdgz)(\nabla^{m}\!B_{m})^{2}
\nonumber\\
& &\,-\,2ih_{_{z}}\left[ 1\,-\,\zdgz\,-\,\frac{1}{3}(1+\ldgl)\right]
(\,u\bar{M}\,-\,\bar{u}M\,)\,\nabla^{m}\!B_{m}
\nonumber\\
& &\,-\,\frac{1}{4}h_{_{z}}\left[ 1\,-\,\zdgz\,-\,\frac{2}{3}(1+\ldgl)\right]
(\,u\bar{M}\,-\,\bar{u}M\,)^{2}
\nonumber\\
& &\,-\,\frac{1}{9}\left[\,3\,+\,(\ldhl-h)(1+\ldgl)\,\right]\bar{M}\!M
\nonumber\\
& &\,-\,\frac{1}{8\dilaton}
\left[\begin{array}{lll}&\,f\,+\,1\,+\,b\dilaton\ln(e^{-k}\bar{u}u/\mu^{6})&\\
&\,+\,\frac{2}{3}(\ldhl+b\dilaton)(1+\ldgl)& \end{array}\right]
(\,u\bar{M}\,+\,\bar{u}M\,)
\nonumber\\
& &\,+\,\frac{1}{4}h_{_{z}}(1-\zdgz)(F_{U}+\bar{F}_{\bar{U}})^{2}
\nonumber\\
& &\,+\,\left\{\begin{array}{lll} 
&\,\frac{1}{8\dilaton}\left[f\,+\,1\,+\,
b\dilaton\ln(e^{-k}\bar{u}u/\mu^{6})\right]& \\
&\,+\,\frac{1}{4\dilaton}(\ldhl+b\dilaton)(1-\zdgz)& \\
&\,-\,\frac{1}{6}h_{_{z}}(1+\ldgl)(\,u\bar{M}\,+\,\bar{u}M\,)&
\end{array}\right\}(F_{U}+\bar{F}_{\bar{U}})
\nonumber\\
& &\,+\,(h+b\dilaton)\sum_{I}\frac{1}{(t^{I}+\bar{t}^{I})^{2}}
\bar{F}_{\bar{T}}^{I}F_{T}^{I}
\nonumber\\
& &\,-\,\frac{1}{16\dilaton^{2}}(\ldhl+h+2b\dilaton)(1+\ldgl)\bar{u}u.
\eea

\bea
\frac{1}{e}\Lag_{\tilde{G}}\,&=&\,\frac{1}{2}\epsilon^{mnpq}
(\,\bar{\psi}_{m}\bar{\sigma}_{n}\!\nabla_{\!p}\psi_{q}\,-\,
\psi_{m}\sigma_{n}\!\nabla_{\!p}\bar{\psi}_{q}\,)
\nonumber\\
& &\,-\,\frac{1}{8\dilaton}\left[\,f\,+\,1\,+\,b\dilaton\ln(e^{-k}\bar{u}u/
\mu^{6})\,\right]\,\bar{u}\,(\psi_{m}\sigma^{mn}\psi_{n})
\nonumber\\
& &\,-\,\frac{1}{8\dilaton}\left[\,f\,+\,1\,+\,b\dilaton\ln(e^{-k}\bar{u}u/
\mu^{6})\,\right]\,u\,(\bar{\psi}_{m}\bar{\sigma}^{mn}\bar{\psi}_{n})
\nonumber\\
& &\,-\,\frac{1}{4}(h+b\dilaton)\sum_{I}\frac{1}{(t^{I}+\bar{t}^{I})}
\epsilon^{mnpq}(\bar{\psi}_{m}\bar{\sigma}_{n}\psi_{p})
(\,\nabla_{\!q}\bar{t}^{I}\,-\,\nabla_{\!q}t^{I}\,)
\nonumber\\
& &\,+\,\frac{i}{4\dilaton}(h+b\dilaton)(1+\ldgl)
(\,\eta^{mn}\eta^{pq}\,-\,\eta^{mq}\eta^{np}\,)
(\bar{\psi}_{m}\bar{\sigma}_{n}\psi_{p})\,\nabla_{\!q}\dilaton
\nonumber\\
& &\,-\,\frac{i}{4}\left[(1-\zdgz)(h+b\dilaton)-1\right]
(\,\eta^{mn}\eta^{pq}\,-\,\eta^{mq}\eta^{np}\,)
(\bar{\psi}_{m}\bar{\sigma}_{n}\psi_{p})\,
\nabla_{\!q}\ln(\bar{u}u)
\nonumber\\
& &\,+\,\frac{1}{4}(h-1+b\dilaton)\,\epsilon^{mnpq}
(\bar{\psi}_{m}\bar{\sigma}_{n}\psi_{p})\,
\nabla_{\!q}\!\ln\!\left(\frac{\bar{u}}{u}\right).
\eea

The bosonic Lagrangian $\Lag_{B}$ contains usual auxiliary fields and the
vector field $B_{m}$ which is dual to an axion. The details of this duality and
the structure of $\Lag_{B}$ will be discussed in the following subsections. The
gravitino Lagrangian $\Lag_{\tilde{G}}$ is in its simplest form. An important 
physical quantity in $\Lag_{\tilde{G}}$ is the gravitino mass 
$m_{_{\tilde{G}}}$ which is the natural order parameter measuring supersymmetry 
breaking. The expression of $m_{_{\tilde{G}}}$ follows directly from 
$\Lag_{\tilde{G}}$:
\be
m_{_{\tilde{G}}}\,=\,\left\langle\,\left|\frac{1}{8\dilaton}\left[\,f\,+\,1\,
+\,b\dilaton\ln(e^{-k}\bar{u}u/\mu^{6})\,\right]u\right|\,\right\rangle.
\ee
Notice that $m_{_{\tilde{G}}}$ contains no moduli $\,T^{I}$ dependence due to 
the Green-Schwarz cancellation mechanism in the linear multiplet formalism of
string models with universal modular anomaly cancellation.
\subsection{Duality Transformation of $B_{m}$}
\hspace{0.8cm}
As pointed out in \cite{Binetruy95,Pillon}, the constraint (1.1) allows us to
interpret the degrees of freedom of $U$ as those of a 3-form supermultiplet, and
the vector field $B_{m}$ is dual to a 3-form $\Gamma^{npq}$. Since 3-form is
dual to 0-form in four dimensions, $B_{m}$ is also dual to a pseudoscalar 
$\axion$. In this subsection, we show explicitly how to rewrite the 
$B_{m}$ part of $\Lag_{B}$ in terms of the dual description using $\axion$.
According to (2.24), the $B_{m}$ terms in $\Lag_{B}$ are:
\bea
\frac{1}{e}\Lag_{B}\;&\ni&\;
+\,\frac{(1+\ldgl)}{4\dilaton^{2}(1-\zdgz)}B^{m}\!B_{m}
\nonumber\\
\;& &\;+\,\frac{i}{2\dilaton}\left[h\,+\,b\dilaton\,-\,
\frac{1}{(1-\zdgz)}\right]\,B^{m}\,
\nabla_{\!m}\!\ln\!\left(\frac{\bar{u}}{u}\right)
\nonumber\\
\;& &\;-\,\frac{i}{2\dilaton}\left[h\,+\,b\dilaton\,-\,
\frac{1}{(1-\zdgz)}\right]\sum_{I}
\frac{(\,\nabla^{m}\bar{t}^{I}\,-\,\nabla^{m}t^{I}\,)}
{(t^{I}+\bar{t}^{I})}B_{m}
\nonumber\\
\;& &\;-\,2ih_{_{z}}\left[ 1\,-\,\zdgz\,-\,\frac{1}{3}(1+\ldgl)\right]
(\,u\bar{M}\,-\,\bar{u}M\,)\,\nabla^{m}\!B_{m}
\nonumber\\
\;& &\;+\,4h_{_{z}}(1-\zdgz)(\nabla^{m}\!B_{m})^{2}.
\eea
They are described by the following generic Lagrangian of $B_{m}$:
\be
\frac{1}{e}\Lag_{B_{m}}\,=\,
\alpha B^{m}\!B_{m}\,+\,\beta\nabla^{m}\!B_{m}\,+\,\zeta^{m}\!B_{m}
\,+\,\tau(\nabla^{m}\!B_{m})^{2}.
\ee
To find the dual description of $\Lag_{B_{m}}$, consider the following
Lagrangian $\Lag_{Dual}$.
\be
\frac{1}{e}\Lag_{Dual}\,=\,
\alpha B^{m}\!B_{m}\,+\,\beta\nabla^{m}\!B_{m}\,+\,\zeta^{m}\!B_{m}
\,+\,\axion\nabla^{m}\!B_{m}\,-\,\frac{1}{4\tau}\axion^{2}.
\ee
In $\Lag_{Dual}$, the auxiliary field $\axion$ acts like a Lagrangian
multiplier, and its equation of motion is:
\be
\axion\,=\,2\tau\nabla^{m}\!B_{m}.
\ee
Therefore, $\Lag_{B_{m}}$ follows directly from $\Lag_{Dual}$ using (2.30).
On the other hand, we can treat the $B_{m}$ in $\Lag_{Dual}$ as auxiliary, and 
write down the equation of motion for $B_{m}$ as follows:
\be
B_{m}\,=\,\frac{1}{2\alpha}
\left(\,\nabla_{m}\axion\,+\,\nabla_{m}\beta\,-\,\zeta_{m}\,\right).
\ee
Eliminating $B_{m}$ from $\Lag_{Dual}$ through (2.31) and then performing a 
field re-definition $\,\axion\,\Rightarrow\,\axion-\beta$, we obtain the 
Lagrangian $\Lag_{\axion}$ of $\axion$:
\be
\frac{1}{e}\Lag_{\axion}\,=\,
-\,\frac{1}{4\alpha}\left(\,\nabla^{m}\!\axion\,-\,\zeta^{m}\,\right)
\left(\,\nabla_{m}\axion\,-\,\zeta_{m}\,\right)
\,-\,\frac{1}{4\tau}\left(\,\axion\,-\,\beta\,\right)^{2}.
\ee
Therefore, $\Lag_{\axion}$ is the dual description of $\Lag_{B_{m}}$ in terms 
of $\axion$ which is interpreted as an axion. Notice that the generation of the
axion mass in $\Lag_{\axion}$ corresponds to the appearance of 
$\,(\nabla^{m}\!B_{m})^{2}\,$ in the dual description. In comparison with 
the model of static gaugino condensation \cite{dilaton}, the model of 
dynamical gaugino condensation has one more axionic degree of freedom $\axion$
that is massive. As will be shown in \mbox{Sect. 4.1}, after integrating out 
the massive axion $\axion$, the axionic contents of the dynamical model are
indeed identical to those of the static model. Therefore, this is consistent
with the fact pointed out in \cite{Binetruy95,Burgess95} that the 
$\,(\nabla^{m}\!B_{m})^{2}\,$ term vanishes in models of static gaugino 
condensation (i.e., $\,h_{_{z}}=0\,$ in (2.27)), and therefore the 
corresponding axionic degree of freedom is massless. 

According to (2.27-28) and (2.32), the $\Lag_{eff}$ defined by (2.23-25) is 
rewritten in the dual description as follows:
\be
\Lag_{eff}\,=\,\Lag_{kin}\,+\,\Lag_{pot}\,+\,\Lag_{\tilde{G}},
\ee
where $\Lag_{kin}$ and $\Lag_{pot}$ refer to the kinetic part and the
non-kinetic part of the bosonic Lagrangian respectively. $\Lag_{\tilde{G}}$
is defined by (2.25). 

\bea
\frac{1}{e}\Lag_{kin}\,&=&\,-\,\frac{1}{2}{\cal R}
\,-\,\frac{1}{4\dilaton^{2}}(h-\ldhl)(1+\ldgl)
\nabla^{m}\!\dilaton\,\nabla_{\!m}\!\dilaton
\nonumber\\
& &\,-\,\frac{(1-\zdgz)}{(1+\ldgl)}\dilaton^{2}
\nabla^{m}\!\axion\,\nabla_{\!m}\!\axion
\,+\,\frac{1}{2\dilaton}\zdhz(1+\ldgl)\,
\nabla^{m}\!\ln(\bar{u}u)\,\nabla_{\!m}\!\dilaton
\nonumber\\
& &\,+\,i\frac{(1-\zdgz)}{(1+\ldgl)}
\left[\,h\,+\,b\dilaton\,-\,\frac{1}{(1-\zdgz)}\,\right]
\dilaton\nabla^{m}\!\axion\,
\nabla_{\!m}\!\ln\!\left(\frac{\bar{u}}{u}\right)
\nonumber\\
& &\,-\,i\frac{(1-\zdgz)}{(1+\ldgl)}
\left[\,h\,+\,b\dilaton\,-\,\frac{1}{(1-\zdgz)}\,\right]
\sum_{I}\frac{(\,\nabla^{m}\bar{t}^{I}\,-\,\nabla^{m}t^{I}\,)}
{(t^{I}+\bar{t}^{I})}\,\dilaton\nabla_{\!m}\!\axion
\nonumber\\
& &\,+\,\frac{1}{4}
\left\{\begin{array}{lll}&\,\zdhz\!\cdot\!\zdgz\frac{(2-\zdgz)}{(1-\zdgz)}& \\
&\,+\,\frac{(1-\zdgz)}{(1+\ldgl)}\left[\,h\,+\,b\dilaton\,-\,
\frac{1}{(1-\zdgz)}\,\right]^{2}& \end{array}\right\}
\frac{1}{\bar{u}^{2}}\nabla^{m}\bar{u}\,\nabla_{\!m}\bar{u}
\nonumber\\
& &\,-\,\frac{1}{2}
\left\{\begin{array}{lll}
&\,\zdhz\left[\,\frac{(2-\zdgz)}{(1-\zdgz)}\,-\,\zdgz\,\right]& \\
&\,+\,\frac{(1-\zdgz)}{(1+\ldgl)}\left[\,h\,+\,b\dilaton\,-\,
\frac{1}{(1-\zdgz)}\,\right]^{2}& \end{array}\right\}
\frac{1}{\bar{u}u}\nabla^{m}\bar{u}\,\nabla_{\!m}u
\nonumber\\
& &\,+\,\frac{1}{4}
\left\{\begin{array}{lll}&\,\zdhz\!\cdot\!\zdgz\frac{(2-\zdgz)}{(1-\zdgz)}& \\
&\,+\,\frac{(1-\zdgz)}{(1+\ldgl)}\left[\,h\,+\,b\dilaton\,-\,
\frac{1}{(1-\zdgz)}\,\right]^{2}& \end{array}\right\}
\frac{1}{u^{2}}\nabla^{m}u\,\nabla_{\!m}u
\nonumber\\
& &\,-\,\frac{1}{2}
\left\{\begin{array}{lll}
&\,\frac{\zdhz}{(1-\zdgz)}& \\
&\,+\,\frac{(1-\zdgz)}{(1+\ldgl)}\left[\,h\,+\,b\dilaton\,-\,
\frac{1}{(1-\zdgz)}\,\right]^{2}& \end{array}\right\}
\sum_{I}\frac{(\,\nabla^{m}\bar{t}^{I}\,-\,\nabla^{m}t^{I}\,)}
{(t^{I}+\bar{t}^{I})}\nabla_{\!m}\!\ln\!\left(\frac{\bar{u}}{u}\right)
\nonumber\\
& &\,+\,\frac{1}{4}
\left\{\begin{array}{lll}
&\,\frac{\zdhz}{(1-\zdgz)}& \\
&\,+\,\frac{(1-\zdgz)}{(1+\ldgl)}\left[\,h\,+\,b\dilaton\,-\,
\frac{1}{(1-\zdgz)}\,\right]^{2}& \end{array}\right\}
\sum_{I,J}\frac{\nabla^{m}\bar{t}^{I}\,\nabla_{\!m}\bar{t}^{J}}
{(t^{I}+\bar{t}^{I})(t^{J}+\bar{t}^{J})}
\nonumber\\
& &\,-\,\frac{1}{2}\sum_{I,J}
\left\{\begin{array}{lll}
&\,2(h+b\dilaton)\delta_{IJ}\,+\,\frac{\zdhz}{(1-\zdgz)}& \\
&\,+\,\frac{(1-\zdgz)}{(1+\ldgl)}\left[\,h\,+\,b\dilaton\,-\,
\frac{1}{(1-\zdgz)}\,\right]^{2}& \end{array}\right\}
\frac{\nabla^{m}\bar{t}^{I}\,\nabla_{\!m}t^{J}}
{(t^{I}+\bar{t}^{I})(t^{J}+\bar{t}^{J})}
\nonumber\\
& &\,+\,\frac{1}{4}
\left\{\begin{array}{lll}
&\,\frac{\zdhz}{(1-\zdgz)}& \\
&\,+\,\frac{(1-\zdgz)}{(1+\ldgl)}\left[\,h\,+\,b\dilaton\,-\,
\frac{1}{(1-\zdgz)}\,\right]^{2}& \end{array}\right\}
\sum_{I,J}\frac{\nabla^{m}t^{I}\,\nabla_{\!m}t^{J}}
{(t^{I}+\bar{t}^{I})(t^{J}+\bar{t}^{J})}.
\eea

\bea
\frac{1}{e}\Lag_{pot}\,&=&\,
\,\frac{h_{_{z}}(1+\ldgl)^{2}}{36(1-\zdgz)}(\,u\bar{M}\,-\,\bar{u}M\,)^{2}
\nonumber\\
& &\,-\,\frac{1}{9}\left[\,3\,+\,(\ldhl-h)(1+\ldgl)\,\right]\bar{M}\!M
\nonumber\\
& &\,-\,\frac{1}{8\dilaton}
\left[\begin{array}{lll}&\,f\,+\,1\,+\,b\dilaton\ln(e^{-k}\bar{u}u/\mu^{6})&\\
&\,+\,\frac{2}{3}(\ldhl+b\dilaton)(1+\ldgl)& \end{array}\right]
(\,u\bar{M}\,+\,\bar{u}M\,)
\nonumber\\
& &\,-\,\frac{i}{4}\left[\,1\,-\,\frac{(1+\ldgl)}{3(1-\zdgz)}\,\right]
\axion (\,u\bar{M}\,-\,\bar{u}M\,)
\nonumber\\
& &\,+\,\frac{1}{4}h_{_{z}}(1-\zdgz)(F_{U}+\bar{F}_{\bar{U}})^{2}
\nonumber\\
& &\,+\,\left\{\begin{array}{lll} 
&\,\frac{1}{8\dilaton}\left[f\,+\,1\,+\,
b\dilaton\ln(e^{-k}\bar{u}u/\mu^{6})\right]& \\
&\,+\,\frac{1}{4\dilaton}(\ldhl+b\dilaton)(1-\zdgz)& \\
&\,-\,\frac{1}{6}h_{_{z}}(1+\ldgl)(\,u\bar{M}\,+\,\bar{u}M\,)&
\end{array}\right\}(F_{U}+\bar{F}_{\bar{U}})
\nonumber\\
& &\,+\,(h+b\dilaton)\sum_{I}\frac{1}{(t^{I}+\bar{t}^{I})^{2}}
\bar{F}_{\bar{T}}^{I}F_{T}^{I}
\nonumber\\
& &\,-\,\frac{1}{16\dilaton^{2}}(\ldhl+h+2b\dilaton)(1+\ldgl)\bar{u}u
\nonumber\\
& &\,-\,\frac{\bar{u}u}{16\zdhz(1-\zdgz)}\,\axion^{2}.
\eea
The $\,b^{a}b_{a}\,$ term has been eliminated by its equation of motion, 
$\,b^{a}=0$, and $\Lag_{kin}$ is in its simplest form. Note that the kinetic
terms of those axionic degrees of freedom $\axion$, $\,i\ln(\bar{u}/u)\,$ and
$\,i(\bar{t}^{I}-t^{I})\,$ are more complicated, which essentially reflects
the non-trivial constraint (1.1) satisfied by $U$ and $\bar{U}$. An important
issue is the structure of $\Lag_{pot}$, and it will be discussed in the next
subsection.

\subsection{Scalar Potential}
\hspace{0.8cm}
It is straightforward to solve the equations of motion for the auxiliary fields
$\,b^{a}$, $\,F_{T}^{I}$, $\,\bar{F}_{\bar{T}}^{I}$, $\,M$, $\,\bar{M}\,$ and 
$(F_{U}+\bar{F}_{\bar{U}})$ respectively as follows:
\bea
b^{a}\,&=&\,0,
\nonumber\\
F_{T}^{I}\,&=&\,0, \;\;\;
\bar{F}_{\bar{T}}^{I}\,=\,0,
\nonumber\\
M\,&=&\,-\,\frac{3}{8\dilaton}\left[\,f\,+\,1\,
+\,b\dilaton\ln(e^{-k}\bar{u}u/\mu^{6})\,\right]u
\,-\,\frac{3iu}{4}\axion,
\nonumber\\
\bar{M}\,&=&\,-\,\frac{3}{8\dilaton}\left[\,f\,+\,1\,
+\,b\dilaton\ln(e^{-k}\bar{u}u/\mu^{6})\,\right]\bar{u}
\,+\,\frac{3i\bar{u}}{4}\axion,
\nonumber\\
(F_{U}+\bar{F}_{\bar{U}})\,&=&\,\frac{(\ldhl-h)}{4\zdhz}
\left[\,f\,+\,1\,+\,b\dilaton\ln(e^{-k}\bar{u}u/\mu^{6})\,\right]
\frac{\bar{u}u}{\dilaton} \nonumber\\
& &\,-\,\frac{(\ldhl+b\dilaton)}{2\zdhz}\!\cdot\!\frac{\bar{u}u}{\dilaton}.
\eea
Note that $\,\langle\,|M|\,\rangle\,=\,3m_{_{\tilde{G}}}\,$ because 
$\,\langle\axion\rangle=0$ always. To obtain the scalar potential, 
the auxiliary fields are eliminated from $\Lag_{eff}$ defined by (2.33), and 
$\Lag_{eff}$ is then rewritten as follows:
\be
\frac{1}{e}\Lag_{eff}\,=\,\frac{1}{e}\Lag_{kin}\,-\,V_{pot}
\,+\,\frac{1}{e}\Lag_{\tilde{G}},
\ee
where $V_{pot}$ is the scalar potential. $\,\Lag_{kin}\,$ and 
$\,\Lag_{\tilde{G}}\,$ are defined by (2.34) and (2.25) respectively. 

\bea
V_{pot}\,&=&\,
\,\frac{1}{16}(\ldhl+h+2b\dilaton)(1+\ldgl)\frac{\bar{u}u}{\dilaton^{2}}
\nonumber\\
& &\,+\,\frac{1}{64\zdhz(1-\zdgz)}
\left\{\begin{array}{lll} 
&\,f\,+\,1\,+\,b\dilaton\ln(e^{-k}\bar{u}u/\mu^{6})& \\
&\,+\,2(\ldhl+b\dilaton)(1-\zdgz)&
\end{array}\right\}^{2}\frac{\bar{u}u}{\dilaton^{2}}
\nonumber\\
& &\,-\,\frac{(2-\ldgl-3\zdgz)}{64(1-\zdgz)}
\left[\,f\,+\,1\,+\,b\dilaton\ln(e^{-k}\bar{u}u/\mu^{6})\,\right]^{2}
\frac{\bar{u}u}{\dilaton^{2}}
\nonumber\\
& &\,+\,\frac{(h-\ldhl-3\zdhz)\bar{u}u}{16\zdhz}\,\axion^{2}.
\eea

Several interesting aspects of $V_{pot}$ can be uncovered. First, 
there is always a trivial vacuum with $\,\left\langle V_{pot}\right\rangle=0\,$
in the specific weak-coupling limit defined as follows:
\be
\dilaton\,\rightarrow\,0,\;\;\;
z\,\rightarrow\,\frac{1}{e^{2}}\dilaton\mu^{6}e^{-1/b\dilaton}
\,\rightarrow\,0,\;\;\;\mbox{and}\;\;\;
g(\dilaton,z),\;f(\dilaton,z)\,\rightarrow\,0.
\ee
Note that quantum corrections to the K\"ahler potential, $g$ and $f$, should
vanish in this limit. As expected, this is consistent with the well-known 
runaway behavior of the dilaton near the weak-coupling limit. 

To proceed further, in the following of this subsection we only study 
$\,V_{pot}\,$ in the $\,z\ll 1\,$ regime. Since a physically interesting model 
of dynamical gaugino condensation should predict a small scale of condensation 
(i.e., $\,\langle z\rangle\ll 1$), there is no loss of generality in this 
choice. Note that in the $\,z\ll 1\,$ regime we have $\,h\approx 1$, 
$\,\ldhl\approx 0$, $\,\zdhz\approx 0\,$ and $\,\zdgz\approx 0\,$ up to small 
corrections that depend on $z$. The structure of $\,V_{pot}\,$ can be analyzed 
as follows: The only axion-dependent term in $\,V_{pot}\,$ is the effective 
axion mass term, the last term in $\,V_{pot}$. In order to avoid a tachyonic
axion, the sign of the effective axion mass term must be positive. Therefore,
the absence of a tachyonic axion requires $\,\zdhz >0$, which is the first 
piece of information about the $\,\bar{U}U$-dependence of the dynamical model. 
Furthermore, $\,\langle\axion\rangle=0$ always, and therefore the last term in 
$\,V_{pot}\,$ is of no significance in discussing the vacuum structure. Because 
of $\,\zdhz >0$, the second term in $\,V_{pot}\,$ is always positive. The signs 
of the first term and the third term in $\,V_{pot}\,$ remain undetermined in
general; however, near the weak-coupling limit the first term is positive and
the third term is negative (which is expected because the third term is the
contribution of auxiliary fields $M$ and $\bar{M}$). Notice that the second 
term in $\,V_{pot}\,$ contains a factor $\,1/\zdhz\,$ ($1/\zdhz\gg 1$), and
therefore it is the dominant contribution to $\,V_{pot}\,$ except near the path
$\,\gamma\,$ defined by $\,\left\{f+1+b\dilaton\ln(e^{-k}\bar{u}u/\mu^{6})
+2(\ldhl+b\dilaton)(1-\zdgz)\right\}=0$. Hence, the vacuum always sits close 
to the path $\,\gamma$. This observation will be essential to the following 
discussion of vacuum structure. 

The second piece of information about the $\,\bar{U}U$-dependence of the 
dynamical model can be obtained as follows. For $\,0<\dilaton<\infty$, the 
first term and the third term in $\,V_{pot}\,$ vanish in the limit 
$\,z\rightarrow 0\,$ generically. If $\,h_{_{z}}\,$ has a pole at $\,z=0$, 
then the second term in $\,V_{pot}\,$ also vanishes for $\,z\rightarrow 0\,$
and $\,0<\dilaton<\infty$. Therefore, for those dynamical models whose 
$\,h_{_{z}}\,$ has a pole at $\,z=0$, there exists a continuous family of 
degenerate vacua (parametrized by $\langle\dilaton\rangle$) with 
$\,\langle z\rangle=0$ (no gaugino condensation), $\,m_{_{\tilde{G}}}=0$ 
(unbroken supersymmetry) and $\,\left\langle V_{pot}\right\rangle=0$. In other
words, in the vicinity of $z=0$ those models always exhibit runaway of $z$ 
toward the degenerate vacua at $z=0$ which do not have the desired physical 
features; whether those models may possess other non-trivial vacuum or not is 
outside the scope of this simple analysis.

On the other hand, the dynamical models whose $\,h_{_{z}}\,$ has no pole at 
$\,z=0\,$ are more interesting. If $\,h_{_{z}}\,$ has no pole at $\,z=0$, 
then $\,V_{pot}\rightarrow\infty\,$ for $\,z\rightarrow 0\,$ and 
$\,0<\dilaton<\infty$. Therefore, for dynamical models whose $\,h_{_{z}}\,$ has
no pole at $\,z=0$, there is no runaway of $z$ toward $\,z=0\,$ except for the
weak-coupling limit (2.39). Furthermore, it implies that gauginos condense 
($\langle z\rangle\neq 0$) if the dilaton is stabilized 
($0<\langle\dilaton\rangle<\infty$). Based on the above observation, it can
actually be shown that supersymmetry is broken ($m_{_{\tilde{G}}}\neq 0$) and 
gauginos condense ($\langle z \rangle\neq 0$) if the dilaton is stabilized: 
As pointed out before, the second term in $\,V_{pot}\,$ is generically the 
dominant contribution. In the following, the second term is rewritten in a 
more instructive form:
\bea
V_{pot}\,&\ni&\,+\,\frac{1}{\zdhz(1-\zdgz)}
\left\{\,M_{_{\tilde{G}}}\,+\,
\frac{1}{4\dilaton}(\ldhl+b\dilaton)(1-\zdgz)|u|\,\right\}^{2},
\eea
where
\be
M_{_{\tilde{G}}}\,\equiv\,\frac{1}{8\dilaton}\left[\,f\,+\,1\,
+\,b\dilaton\ln(e^{-k}\bar{u}u/\mu^{6})\,\right]\!\cdot\!|u|.
\ee
The gravitino mass is related to $M_{_{\tilde{G}}}$ by 
$\,m_{_{\tilde{G}}}\,=\,\left\langle\left|M_{_{\tilde{G}}}\right|\right\rangle$.
If $\,0<\langle\dilaton\rangle<\infty$, then $\,\langle z\rangle\neq 0\,$ and
the second term should vanish at the vacuum (up to small corrections of order 
$\langle z\rangle$). That is, $\,m_{_{\tilde{G}}}=
\left\langle\left|M_{_{\tilde{G}}}\right|\right\rangle\approx
\left\langle\frac{1}{4\dilaton}(\ldhl+b\dilaton)(1-\zdgz)|u|\right\rangle
\approx\frac{1}{4}b\langle|u|\rangle\neq 0\,$ (up to small corrections 
of order $\langle\bar{u}u\rangle$). Therefore, for dynamical models
whose $\,h_{_{z}}\,$ has no pole at $\,z=0$, {\em supersymmetry is broken and 
gauginos condense if the dilaton is stabilized}. The same conclusion has also 
been established in the study of static gaugino condensation \cite{dilaton}, 
which will be briefly reviewed in the next section.

As pointed out before, kinetic terms of the gaugino condensate $U$
naturally arise from field-theoretical loop corrections as well as from 
classical string corrections. As will be discussed in \mbox{Sect. 4} 
these kinetic terms are S-duality invariant \cite{zumino} and correspond to 
corrections $\,\bar{U}U/V^{2}$, $\,\left(\bar{U}U/V^{2}\right)^{2},\cdots$ 
to the K\"ahler potential. This interesting class of S-dual dynamical models 
obviously belongs to the dynamical models whose $\,h_{_{z}}\,$ has no pole at 
$\,z=0$, and therefore it has the nice features established in the previous 
paragraph. In \mbox{Sect. 4}, S-dual dynamical models as well as the issue of
dilaton stabilization will be studied in detail. 
\section{Review of Static Gaugino Condensation} 
\setcounter{equation}{0}
\hspace{0.8cm}
Those features of static condensation \cite{dilaton} which are essential to the
study of S-dual dynamical models in \mbox{Sect. 4} are briefly reviewed here.
Considering the same string models as those in \mbox{Sect. 2}, we write the
generic model of a static E$_{8}$ gaugino condensate as follows:
\bea
K\,&=&\,\ln V\,+\,g(V)\,+\,G, \nonumber \\
\Lag_{eff}\,&=&\,\superint\,E\,\left\{\,\left(\,-2\,+\,f(V)\,\right)\,+\, 
bVG\,+\,bV\ln(e^{-K}\bar{U}U/\mu^{6})\,\right\},\hspace{1.5cm}
\\
& &V\frac{\diff g(V)}{\diff V}\,=\,
-V\frac{\diff f(V)}{\diff V}\,+\,f,
\\
& &g(V=0)\,=\,0 \;\;\;\mbox{and}\;\;\; f(V=0)\,=\,0. 
\eea 
The K\"ahler potential depend only on $V$, and the condensate $U$ is therefore
static. $g(V)$ and $f(V)$ represent quantum corrections to the K\"ahler 
potential. (3.2) guarantees the correct normalization of the Einstein term. The
boundary condition (3.3) in the weak-coupling limit is fixed by the tree-level 
K\"{a}hler potential. Unlike the partial differential equation (2.20), (3.2) is
an ordinary differential equation, and therefore $g(V)$ is unambiguously
related to $f(V)$. Two important physical quantities of the static model are 
the gaugino condensate and the gravitino mass:
\bea
\bar{u}u\,&=&\,\frac{1}{e^{2}}\dilaton\mu^{6}
e^{g\,-\,({f+1})/{b\dilaton}}. \\
m_{_{\tilde{G}}}\,&=&\,\frac{\,1\,}{\,4\,}b\,
\langle\,|u|\,\rangle.
\eea
They imply that {\em supersymmetry is broken and gauginos condense if the 
dilaton is stabilized}. These three issues are unified elegantly. Furthermore,
supersymmetry is broken in the dilaton direction rather than in the direction
of modulus $T^{I}$. The generic expression of scalar potential, which 
depends only on $\dilaton$, is:
\be
V_{pot}\,=\,\frac{1}{16e^{2}\dilaton}\left\{\,
(1+f-\ldfl)(1+b\dilaton)^{2}\,-\,3b^{2}\dilaton^{2}\,\right\}
\mu^{6}e^{g\,-\,({f+1})/{b\dilaton}}.
\ee

In order to appreciate the significance of quantum corrections $g(V)$ and 
$f(V)$, a simple model with tree-level K\"ahler potential (i.e., (3.1) with 
$g(V)$=$f(V)$=0) is considered, and its scalar potential $\,V_{pot}\,$ is shown
in \mbox{Fig. 1-A}. Its $\,V_{pot}\,$ is unbounded from below in the
strong-coupling limit $\,\dilaton\rightarrow\infty$, which is caused by a term
of two-loop order, $-2b^{2}\dilaton^{2}$, in $\,V_{pot}$. This unboundedness
simply reflects that (non-perturbative) quantum corrections, $g(V)$ and $f(V)$, 
to the K\"ahler potential should not be ignored, especially in the 
strong-coupling regime. It can be shown that the necessary and sufficient 
condition for $V_{pot}$ to be bounded from below is:
\be
f-\ldfl\,\geq\,2\;\;\;\;\;\;\mbox{for}\;\;\;\dilaton\,\rightarrow\,\infty.
\ee
(3.7) can also be interpreted as the necessary condition for the dilaton to be
stabilized. Furthermore, it has been argued in detail \cite{dilaton} that
non-perturbative quantum corrections to the K\"ahler potential may naturally 
stabilize the dilaton if (3.7) is satisfied. A nice realization of that 
argument is shown in \mbox{Fig. 1-B}, where the dilaton is stabilized and 
supersymmetry is broken with (fine-tuned) vanishing cosmological constant. 

As the conclusion of this section, we comment on the meaning of the quantum 
corrections, $\,g(V)\,$ and $\,f(V)$, to the K\"ahler potential. Consider the
unconfined string effective Lagrangian at the string scale $M_{S}$:
\bea
K\,&=&\,\ln L\,+\,g(L)\,+\,G, \nonumber \\
\Lag_{eff}\,&=&\,\superint\,E\,\left\{\,\left(\,-2\,+\,f(L)\,\right)\,+\, 
bLG\,\right\}, \\
{\cal W}^{\alpha}{\cal W}_{\alpha}\,&=&\,-(\DbDb-8R)L, \nonumber\\
{\cal W}_{\dot{\alpha}}{\cal W}^{\dot{\alpha}}\,&=&\,-(\DaDa-8R^{\dagger})L,
\nonumber
\eea 
whose confined theory corresponds to (3.1). It is straightforward to compute 
the gauge coupling at the string scale, $g(M_{S})$, defined by (3.8) as follows:
\be
g^{2}(M_{S})\,=\,\left\langle\,\frac{2\dilaton}{(\,1\,+\,f\,)}\,\right\rangle.
\ee
According to (3.9), the $\,\bar{u}u$'s exponential dependence on $g^{2}(M_{S})$ 
in (3.4) is consistent with the well-known analysis using renormalization 
group. (3.9) is also consistent with the interpretation of $g^{2}(M_{S})$ in 
the chiral formalism of (3.8)\footnote{The chiral formalism of (3.8) is 
obtained by performing a duality transformation \cite{Binetruy91,Gaillard92}.}:
In the chiral formalism, we always have 
$\,g^{2}(M_{S})\,=\,\langle\,2/(s+\bar{s})\,\rangle$, where $S$ is the dilaton 
chiral superfield and $s=S\lowest$. On the other hand, it has been shown 
\cite{Gaillard92} that $\,1/(S+\bar{S})\,$ corresponds to $\,L/(1+f)\,$ through
a duality transformation from the linear multiplet formalism of (3.8) to the
chiral formalism of (3.8). Therefore, the interpretations of $g^{2}(M_{S})$ in
both formalisms are consistent with each other. In the absence of $\,g(L)\,$ 
and $\,f(L)$, we have the usual relation 
$\,g^{2}(M_{S})\,=\,2\langle\dilaton\rangle\,$ at the string scale 
\cite{Gaillard92}. Therefore, the $\,1/(1+f)\,$ factor in (3.9) is naturally 
interpreted as the renormalization of $g^{2}(M_{S})$ by effects above the 
string scale; $\,g(L)\,$ and $\,f(L)\,$ are then interpreted as  
stringy corrections to the K\"ahler potential. 

The above observation implies that the non-perturbative corrections, $\,g(V)\,$
and $\,f(V)$, to the K\"ahler potential of (3.1) should be interpreted as 
stringy non-perturbative corrections. In this interpretation, it is actually 
stringy non-perturbative effects that stabilize the dilaton and allow
dynamical supersymmetry breaking via the field-theoretical non-perturbative 
effect of gaugino condensation. Furthermore, (3.7) is now interpreted as the 
necessary condition for stringy non-perturbative effects to stabilize the
dilaton.\footnote{In the presence of significant stringy non-perturbative 
effects, (3.9) could have implications for gauge coupling unification. This 
is considered in the study of multi-gaugino and matter condensation 
\cite{multiple}.} As we shall see in the next section, stringy non-perturbative 
effects also play the same crucial role in generic models of dynamical gaugino 
condensation.
\section{S-Dual Model of Dynamical Gaugino Condensation} 
\setcounter{equation}{0}
\hspace{0.8cm}
As has been discussed in \mbox{Sect. 2.3}, one of the motivations for studying
models of dynamical gaugino condensation is the observation that kinetic terms 
of the gaugino condensate naturally arise from field-theoretical loop 
corrections as well as from classical string corrections. For example, the
relevant field-theoretical one-loop correction has been computed using chiral 
formalism \cite{Jain96,sduality}: 
\be
\Lag_{one-loop}\;\ni\;\frac{N_{G}}{128\pi^{2}}
\superint\,E\,(S+\bar{S})^{2}\,({\cal W}^{\alpha}{\cal W}_{\alpha})\,
({\cal W}_{\dot{\alpha}}{\cal W}^{\dot{\alpha}})\,\ln\Lambda^{2},
\ee
where $\Lambda$ is the effective cut-off and $N_{G}$ is the number of gauge 
degrees of freedom. Therefore, the confined theory using linear multiplet
formalism should contain a term which corresponds to (4.1):
\be
\Lag_{eff}\;\ni\;\superint\,E\,\frac{\bar{U}U}{V^{2}},
\ee
as well as higher-order corrections $\,\left(\bar{U}U/V^{2}\right)^{2}$,
$\,\left(\bar{U}U/V^{2}\right)^{3},\cdots .\,$ These D terms are corrections to
the K\"ahler potential, and will generate the kinetic terms for gaugino
condensate $U$. An interesting interpretation of these corrections is that they
are S-duality invariant in the sense defined by Gaillard and Zumino 
\cite{zumino}. This S-duality, which is an SL(2,R) symmetry among elementary
fields, is a symmetry of the equations of motion only of the
dilaton-gauge-gravity sector in the limit of vanishing gauge couplings. The
implication of this S-duality for gaugino condensation has recently been 
studied in \cite{sduality} using the chiral formalism. 

Motivated by the above observation, we consider in this section models of 
dynamical gaugino condensation where the kinetic terms for gaugino condensate 
arise from the S-dual loop corrections defined by (4.2). More precisely, we
consider the following dynamical model:
\bea
K\,&=&\,\ln V\,+\,g(V,X)\,+\,G, 
\nonumber \\
\Lag_{eff}\,&=&\,
\superint\,E\,\left\{\,\left(\,-2\,+\,f(V,X)\,\right)\,+\,bVG\,
+\,bV\ln(e^{-K}\bar{U}U/\mu^{6})\,\right\},\hspace{1.5cm}
\eea
\be
\left(2+X\frac{\partial f}{\partial X}\right)
\left(1-V\frac{\partial g}{\partial V}\right)=
\left(2-X\frac{\partial g}{\partial X}\right)
\left(1-f+V\frac{\partial f}{\partial V}\right).
\ee
For convenience, we have written the S-dual combination
$\,(\bar{U}U)^{\frac{1}{2}}/V\,$ as a vector superfield $X$, and
therefore its lowest component $\,x\,=\,X\lowest\,$ is 
$x\,=\,(\bar{u}u)^{\frac{1}{2}}/\dilaton\,=\,\sqrt{z}/\dilaton$.
Eq.(4.4) guarantees the correct normalization of the Einstein term. 
$\,g(V,X)\,$ and $\,f(V,X)\,$ satisfy the boundary condition in the
weak-coupling limit defined by (2.39). We also assume that $\,g(V,X)\,$ and 
$\,f(V,X)\,$ have the following power-series representations\footnote{It should
be noted that one can actually start with a more generic dynamical model by
considering more generic $\,g(V,X)\,$ and $\,f(V,X)\,$, and the discussions of
\mbox{Sect. 4} remain valid.} in terms of $X^{2}$:
\bea
g(V,X)\,&\equiv&\,g^{(0)}(V)\,+\,g^{(1)}(V)\!\cdot\!X^{2}
\,+\,g^{(2)}(V)\!\cdot\!X^{4}\,+\,\cdots.
\nonumber\\
f(V,X)\,&\equiv&\,f^{(0)}(V)\,+\,f^{(1)}(V)\!\cdot\!X^{2}
\,+\,f^{(2)}(V)\!\cdot\!X^{4}\,+\,\cdots.
\eea
Furthermore, $\,g^{(n)}(V)\,$ and $\,f^{(n)}(V)\,$ ($n\geq 0$) are assumed to 
be arbitrary but bounded here. The interpretation of each term in (4.5) is 
obvious: As has been discussed at the end of \mbox{Sect. 3}, $\,g^{(0)}(V)\,$ 
and $\,f^{(0)}(V)\,$ are stringy non-perturbative corrections to the K\"ahler 
potential. $\,g^{(n)}(V)\!\cdot\!X^{2n}\,$ and 
\mbox{$\,f^{(n)}(V)\!\cdot\!X^{2n}\,$} ($n\geq 1$) are S-dual loop corrections 
to the K\"ahler potential in the presence of stringy non-perturbative effects. 

It is also more convenient to use the coordinates $\,(\,\dilaton,\,x\,)\,$ 
instead of $\,(\,\dilaton,\,z\,)\,$ for the field configuration space. The 
component field expressions constructed in \mbox{Sect. 2} can easily be 
rewritten in the new coordinates $\,(\,\dilaton,\,x\,)\,$ according to the 
following rules:
\bea
\ldgl\;&\rightarrow&\;\ldgl\,-\,\xdgx \nonumber\\
\zdgz\;&\rightarrow&\;\frac{1}{2}\xdgx,
\eea
where
\be
g_{_{\dilaton}}\,\equiv\,\frac{\partial g(\dilaton,x)}{\partial \dilaton},
\;\;\;g_{_{x}}\,\equiv\,\frac{\partial g(\dilaton,x)}{\partial x}
\ee
on the right-hand side of (4.6) are to be understood as partial derivatives in 
the coordinates $\,(\,\dilaton,\,x\,)$. The scalar potential of this generic 
model follows directly from (2.38):
\bea
V_{pot}\,&=&\,\frac{1}{16}(\,1\,+\,\ldgl\,-\,\xdgx\,)
(\,h\,+\,\ldhl\,-\,\xdhx\,+\,2b\dilaton\,)\,x^{2}
\nonumber\\
& &\,+\,\frac{1}{16\xdhx(\,2\,-\,\xdgx\,)}
\left\{\begin{array}{lll} 
&\,f\,+\,1\,+\,b\dilaton\ln(e^{-k}\bar{u}u/\mu^{6})& \\
&\,+\,(\,2\,-\,\xdgx\,)(\,\ldhl\,-\,\xdhx\,+\,b\dilaton\,)&
\end{array}\right\}^{2}x^{2}
\nonumber\\
& &\,-\,\frac{(\,4\,-\,2\ldgl\,-\,\xdgx\,)}{64(\,2\,-\,\xdgx\,)}
\left[\,f\,+\,1\,+\,b\dilaton\ln(e^{-k}\bar{u}u/\mu^{6})\,\right]^{2}x^{2}
\nonumber\\
& &\,+\,\frac{(\,2h\,-\,2\ldhl\,-\,\xdhx\,)\bar{u}u}{16\xdhx}\,\axion^{2}.
\eea
The kinetic terms also follow directly from (2.34). The absence of a tachyonic 
axion requires $\,\xdhx>0$. 

As discussed in \mbox{Sect. 2.4}, the S-dual dynamical model considered here 
belongs to the dynamical models whose $\,h_{_{z}}\,$ has no pole at $\,z=0$; 
part of its vacuum structure has already been analyzed in \mbox{Sect. 2.4}. It 
is concluded that supersymmetry is broken if the dilaton is stabilized. 
Therefore, we will focus on the issue of dilaton stabilization in the following 
subsection.
\subsection{Low-Energy Limit of Dynamical Gaugino Condensation}
\hspace{0.8cm}
Since a physically interesting model of dynamical gaugino condensation should 
predict a small scale of condensation (i.e., $\,\langle x\rangle\ll 1$), it is
clear from (4.8) that generally the condensate $x$ and the axion $\axion$ are
much heavier than the other fields, and therefore should be integrated out. It
is straightforward to integrate out $\axion$ and $x$ through their equations of
motion: The equation of motion for $\axion$ is $\,\axion=0$. The equation of 
motion for $x$ is
\be
f\,+\,1\,+\,b\dilaton\ln(e^{-k}\bar{u}u/\mu^{6})\,+\,(\,2\,-\,\xdgx\,)
(\,\ldhl\,-\,\xdhx\,+\,b\dilaton\,)\,=\,0\,+\,\mbox{O}\left(x^2\right).
\ee
(4.9) can be re-written in a more instructive form:
\be
x^{2}\;=\;\frac{\mu^{6}}{e^{2}\dilaton}\,
e^{g^{(0)}\,-\,\left({f^{(0)}+1}\right)/{b\dilaton}}\,+\,
\mbox{O}\left(x^4\right),
\ee
where we have used the fact that \mbox{$\,g\approx g^{(0)}$}, 
\mbox{$\,f\approx f^{(0)}$}, \mbox{$\,h\approx 1$}, 
\mbox{$\,\ldgl\approx \dilaton g^{(0)}_{_{\dilaton}}$}, 
\mbox{$\,\ldfl\approx \dilaton f^{(0)}_{_{\dilaton}}$}, 
\mbox{$\,\ldhl\approx 0$}, \mbox{$\,\xdgx\approx 0$}, 
\mbox{$\,\xdfx\approx 0\,$} and \mbox{$\,\xdhx\approx 0\,$} up to corrections 
of order $\mbox{O}\left(x^2\right)$. The (bosonic) effective Lagrangian, 
$\,\Lag_{eff}\,=\,\Lag_{kin}\,-\,eV_{pot}$, of the dynamical model (4.3-5) 
after integrating out $\axion$ and $x$ is as follows:

\bea
\frac{1}{e}\Lag_{kin}\,&=&\,-\,\frac{1}{2}{\cal R}
\,-\,\frac{1}{4\dilaton^{2}}\left(1+\dilaton g^{(0)}_{_{\dilaton}}\right)
\nabla^{m}\!\dilaton\,\nabla_{\!m}\!\dilaton
\nonumber\\
& &\,-\,(1+b\dilaton)\sum_{I}\frac{1}{(t^{I}+\bar{t}^{I})^{2}}
\nabla^{m}\bar{t}^{I}\,\nabla_{\!m}t^{I}
\,+\,\frac{1}{4\dilaton^{2}}\left(1+\dilaton g^{(0)}_{_{\dilaton}}\right)
\tilde{B}^{m}\!\tilde{B}_{m}
\nonumber\\
& &\,+\,\mbox{O}\left(x^2\right),
\eea
where
\bea
\tilde{B}_{m}\,&\equiv&
\,-\,i\frac{b\dilaton^{2}}{\left(1+\dilaton g^{(0)}_{_{\dilaton}}\right)}
\nabla_{\!m}\!\ln(\frac{\bar{u}}{u}) \nonumber\\
& &\,+\,i\frac{b\dilaton^{2}}{\left(1+\dilaton g^{(0)}_{_{\dilaton}}\right)}
\sum_{I}\frac{1}{(t^{I}+\bar{t}^{I})}
(\,\nabla_{\!m}\bar{t}^{I}\,-\,\nabla_{\!m}t^{I}\,).
\eea

\bea
V_{pot}\,&=&\,\frac{1}{16e^{2}\dilaton}\left\{\,
\left(1+f^{(0)}-\dilaton f^{(0)}_{_{\dilaton}}\right)
\left(1+b\dilaton\right)^{2}\,-\,3b^{2}\dilaton^{2}\,\right\}\mu^{6}
e^{g^{(0)}\,-\,\left({f^{(0)}+1}\right)/{b\dilaton}} 
\nonumber\\
& &\,+\,\mbox{O}\left(x^4\right).
\eea
Furthermore, (4.4) leads to $\,\dilaton g^{(0)}_{_{\dilaton}}=
f^{(0)}-\dilaton f^{(0)}_{_{\dilaton}}\,$ to the lowest order in $x^2$.

In comparison with the static model of gaugino condensation \cite{dilaton}, 
it is clear that the low-energy effective Lagrangian of the dynamical model 
are identical to the Lagrangian of the static model to the lowest order in 
$x^2$. Note that, in (4.13), the $\mbox{O}\left(x^4\right)$ terms do not depend 
on the axionic degrees of freedom (i.e., $\,i\ln(\bar{u}/u)\,$ and 
$\,i(\bar{t}^{I}-t^{I})$), and therefore these axions remain massless, as
expected.\footnote{On the other hand, these axionic degrees of freedom 
naturally acquire masses in scenarios of multiple gaugino condensation 
\cite{multiple}.} According to the equation of motion for $x$, (4.10), 
$\,x^{2}\ll 1\,$ actually holds for any value of $\dilaton$. It implies that
only the lowest-order terms (in $x^2$) of (4.11) and (4.13) are important, and 
therefore the static model of gaugino condensation is indeed the appropriate 
low-energy effective description of the dynamical model. The above observation 
implies that the necessary and sufficient condition for $V_{pot}$ of the 
dynamical model to be bounded from below is exactly the same as that of the 
static model:
\be
f^{(0)}-\dilaton f^{(0)}_{_{\dilaton}}\,\geq\,2\;\;\;\;\;\;
\mbox{for}\;\;\;\dilaton\,\rightarrow\,\infty,
\ee
which depends only on stringy non-perturbative effects $g^{(0)}$ and 
$f^{(0)}$. (4.14) does not depend on the details of S-dual loop corrections, 
and therefore it holds for generic S-dual dynamical models. As discussed in 
\mbox{Sect. 3}, (4.14) can also be interpreted as the necessary condition for 
the dilaton to be stabilized. The above analysis shows that it is indeed 
stringy non-perturbative effects that stabilize the dilaton and allow 
supersymmetry breaking via gaugino condensation. 
\subsection{Solving for Dynamical Gaugino Condensation}
\hspace{0.8cm}
In the previous subsection, the dynamical model of gaugino condensation is 
analyzed through its low-energy effective Lagrangian. One can also analyze 
the dynamical model directly, and obtain the same conclusion. Here, we would
like to present a typical example of dynamical gaugino condensation as a
concrete supplement to the analysis of \mbox{Sect. 4.1}. Solving for dynamical 
gaugino condensation is generically difficult due to the partial differential 
equation, (2.20) or (4.4), which guarantees the correct normalization of the 
Einstein term. On the other hand, only those solutions of (2.20) which are of 
physical interest deserve study. Therefore, in the following we show 
explicitly how to construct the solution for the interesting S-dual model of 
dynamical gaugino condensation defined by (4.3-5). In order to simplify the 
presentation but leave the generality of our conclusion unaffected, we choose 
a specific form for $f(V,X)$ in the following discussion: 
\mbox{$\,f(V,X)\,=\,f^{(0)}(V)\,+\,\ep X^{2}$,} where $\ep$ is a constant and 
$|\ep|$ is in principle a small number because $X$-dependent terms arise from 
loop corrections. In this restricted solution space, (4.4) together with the 
boundary condition (2.39) can be re-expressed as an infinite number of 
ordinary differential equations with appropriate boundary conditions 
(evaluated at \mbox{$\theta =\bar{\theta}=0$}) as follows:
\bea
\dilaton g^{(0)}_{_{\dilaton}}&=&f^{(0)}-\dilaton f^{(0)}_{_{\dilaton}}.
\nonumber\\
\dilaton g^{(1)}_{_{\dilaton}}-
\left(1-f^{(0)}+\dilaton f^{(0)}_{_{\dilaton}}\right)g^{(1)}&=&
-\ep\!\cdot\!\dilaton g^{(0)}_{_{\dilaton}}+2\ep.
\nonumber\\
\dilaton g^{(n)}_{_{\dilaton}}-
n\left(1-f^{(0)}+\dilaton f^{(0)}_{_{\dilaton}}\right)g^{(n)}&=&
-\ep\!\cdot\!\dilaton g^{(n-1)}_{_{\dilaton}}-\ep(n-1)g^{(n-1)},
\nonumber\\
& &\mbox{for}\;\;\;n\geq 2.
\eea
The associated boundary conditions in the weak-coupling limit are:
\bea
& &g^{(0)}(\dilaton=0)\,=\,0,\;\;\;f^{(0)}(\dilaton=0)\,=\,0,
\nonumber\\
& &g^{(1)}(\dilaton=0)\,=\,-2\ep,
\nonumber\\
& &g^{(n)}(\dilaton=0)\,=\,-\,\frac{2}{n}\,\ep^{n}
\;\;\;\;\;\mbox{for}\;\;\;n\geq 2.
\eea
Therefore, $\,g(V,X)\,$ is unambiguously\footnote{In fact, there is one free
parameter $\beta$ involved due to the fact that 
$\,g^{(n)}_{_{\dilaton}}(\dilaton=0)\,$ is not well-defined in (4.15); this 
ambiguity can be parametrized by 
$\,g^{(n)}_{_{\dilaton}}(\dilaton=0)\,=\,n\ep^{n-1}\beta$. We take $\beta=0$
here.} related to $\,f(V,X)\,$ in this interesting solution space. 

First, notice that the boundedness of $g^{(n)}$ and $f^{(n)}$ can be 
guaranteed if (4.14) is satisfied. Therefore, the solution defined by 
(4.15-16)\footnote{The generalization to generic $f(V,X)$ is straightforward.} 
exists at least for viable dynamical models in the sense of (4.14). Secondly, 
$g^{(n)}$ is suppressed by a small factor $|\ep|^{n}$, which is obvious from 
(4.15-16). Therefore, the solution defined by (4.15-16) converges for 
$\,x^{2}<\mbox{O}\left(1/\ep\right)$. Since a physically interesting model 
of gaugino condensation should predict a small scale of condensation (i.e., 
$\,\langle x^{2}\rangle\ll 1$), this solution does cover the regime of physical
interest.\footnote{This solution can in principle be extended into the 
$\,x^{2}>\mbox{O}\left(1/\ep\right)\,$ regime using the method of 
characteristics.} 

(4.14) is the necessary condition for stringy non-perturbative effects to
stabilize the dilaton. By looking into the details of the scalar potential, it
can also be argued \cite{dilaton} that stringy non-perturbative corrections to 
the K\"ahler potential may naturally stabilize the dilaton if (4.14) is 
satisfied. In the following, the solution defined by (4.15-16) is used to 
construct a typical realization of this argument. Furthermore, as illustrated 
in \mbox{Sect. 1}, it is the typical feature of this example rather than the 
specific form of $g(V,X)$ and $f(V,X)$ assumed in this example that 
we want to emphasize. In \mbox{Fig. 2}, the scalar potential $\,V_{pot}\,$ is 
plotted versus $\,\dilaton\,$ and $\,x\,$ for an example with 
\mbox{$\,f(V,X)=f^{(0)}(V)+\ep X^{2}\,$} and 
\mbox{$\,f^{(0)}(V)=A\!\cdot\!e^{-B/V}$}. There is a non-trivial vacuum with 
the dilaton stabilized at $\,\langle\dilaton\rangle=0.52$, $\,x\,$ stabilized 
at $\,\langle x\rangle=\langle\sqrt{\bar{u}u}/\dilaton\rangle=0.0024$, and
(fine-tuned) vanishing vacuum energy $\,\left\langle V_{pot}\right\rangle=0$. 
Supersymmetry is broken at the vacuum and the gravitino mass 
$\,m_{_{\tilde{G}}}=4\times 10^{-4}\,$ in reduced Planck units. 
To uncover more details of dilaton 
stabilization in \mbox{Fig. 2}, a cross section of $\,V_{pot}\,$ is 
presented in \mbox{Fig. 4}. More precisely, with the value of $\dilaton\,$ 
fixed, $\,V_{pot}\,$ is minimized only with respect to $\,x$; the location of 
this minimum is denoted as \mbox{($\dilaton$, $x_{min}(\dilaton)$).} The path 
defined by \mbox{($\dilaton$, $x_{min}(\dilaton)$)} is shown in \mbox{Fig. 3}. 
The cross section of $\,V_{pot}\,$ is obtained by making a cut along 
\mbox{($\dilaton$, $x_{min}(\dilaton)$)}; that is, the cross section of 
$\,V_{pot}\,$ is defined as \mbox{$\,V'_{pot}(\dilaton)\equiv 
V_{pot}\left(\dilaton,x_{min}(\dilaton)\right)$}. \mbox{Fig. 4} shows that the 
dilaton is indeed stabilized at \mbox{$\,\langle\dilaton\rangle=0.52$.} 
Therefore, we have presented a concrete example with stabilized dilaton, broken
supersymmetry, and (fine-tuned) vanishing cosmological constant. One can also 
consider the stringy non-perturbative effect conjectured by \cite{Shenker90}, 
and the generic feature is similar to that of \mbox{Fig. 2}. As pointed out in 
\mbox{Sect. 1}, this is in contrast with condensate models studied previously 
\cite{Dine85,racetrack,constant} which either need the assistance of an
additional source of supersymmetry breaking or have a large and negative 
cosmological constant. 
\section{Concluding Remarks}
\hspace{0.8cm}\setcounter{equation}{0}
This paper begins with two generic questions in the context of dynamical 
gaugino condensation: What is the generic condition for the dilaton to be 
stabilized? Is supersymmetry broken if the dilaton is stabilized? First, it is 
emphasized that the linear multiplet formalism of gaugino condensation is the
framework in which these two questions can be defined and answered more 
appropriately. Secondly, the field component Lagrangian for the linear multiplet
formalism of generic dynamical gaugino condensation is constructed as the
grounds of this study; it may also be useful to future studies.

The second question can be answered in a very generic context: by analyzing the 
vacuum structure of generic dynamical models, it is found that, for dynamical 
models whose $\,h_{_{z}}\,$ has no pole at $\,z=0$, supersymmetry is broken if 
the dilaton is stabilized. In particular, a class of well-motivated models, the
S-dual model of dynamical gaugino condensation, does belong to this category. 

As for the first question, it is shown that the low-energy limit of dynamical
gaugino condensation is appropriately described by static gaugino condensation. 
An interesting necessary condition (4.14) for the dilaton to be stabilized, 
which was first derived in the study of static gaugino condensation, is then 
shown to hold for generic S-dual models of dynamical gaugino condensation.
Furthermore, the analysis of (4.14) shows that it is stringy non-perturbative 
effects that stabilize the dilaton and allow dynamical supersymmetry breaking 
via the field-theoretical non-perturbative effect of gaugino condensation. We 
also present a concrete example where the dilaton is stabilized and 
supersymmetry is broken.

For the string models considered here, supersymmetry is broken in the dilaton 
direction rather than in the direction of modulus $T^{I}$. However, the 
hierarchy between the Planck scale and the gravitino mass generated by the 
confined E$_{8}$ hidden sector is insufficient to account for the observed 
scale of electroweak symmetry breaking. This is simply due to the large gauge 
content of gauge group E$_{8}$. In realistic string models, the hidden-sector 
gauge group is in general a product group, and the gauge content of each 
non-abelian subgroup is smaller than that of E$_{8}$. Therefore, the hierarchy 
generated by realistic string models could be sufficient to explain the scale 
of electroweak symmetry breaking. On the other hand, the generalization of the 
current study to realistic string models is very non-trivial since multiple 
gaugino condensation as well as hidden matter condensation occurs in generic 
hidden sectors. Furthermore, the cancellation of modular anomaly is also a very
important issue. These issues together with the stabilization of the moduli are 
considered in \cite{multiple}. 
\section*{Acknowledgements}
\hspace{0.8cm}
I sincerely thank Pierre Bin\'{e}truy and Mary K. Gaillard for their important
contributions. This work was supported in part by the Director, Office of 
Energy Research, Office of High Energy and Nuclear Physics, Division 
of High Energy Physics of the U.S. Department of Energy under Contract 
DE-AC03-76SF00098 and in part by the National Science Foundation under 
grant PHY-95-14797.
\pagebreak

\pagebreak
\clearpage
\hspace{1.6in} FIGURE CAPTIONS

\vskip 0.5in
\mbox{Fig. 1-A:} The scalar potential $\,V_{pot}\,$ (in reduced Planck units)
is plotted versus the dilaton $\,\dilaton\,$ without non-perturbative 
corrections to the K\"ahler potential.

\vskip 0.5in
\mbox{Fig. 1-B:} The scalar potential $\,V_{pot}\,$ (in reduced Planck units)
is plotted versus the dilaton $\,\dilaton\,$ with appropriate non-perturbative 
corrections to the K\"ahler potential.

\vskip 0.5in
\mbox{Fig. 2:} The scalar potential $\,V_{pot}\,$ (in reduced Planck units)
is plotted versus $\,\dilaton\,$ and $\,x$. $\,A=6.8$, $\,B=1$, $\,\ep=-0.1\,$ 
and $\,\mu$=1. (The rippled surface of $V_{pot}$ is simply due to 
discretization of $\dilaton$-axis.)

\vskip 0.5in
\mbox{Fig. 3:} $\,x_{min}(\dilaton)\,$ is plotted versus $\,\dilaton\,$ for 
\mbox{Fig. 2.}

\vskip 0.5in
\mbox{Fig. 4:} The cross section of the scalar potential, 
$V'_{pot}(\dilaton)\equiv V_{pot}\left(\dilaton,x_{min}(\dilaton)\right)\,$ 
(in reduced Planck units), is plotted versus $\,\dilaton\,$ for \mbox{Fig. 2.}

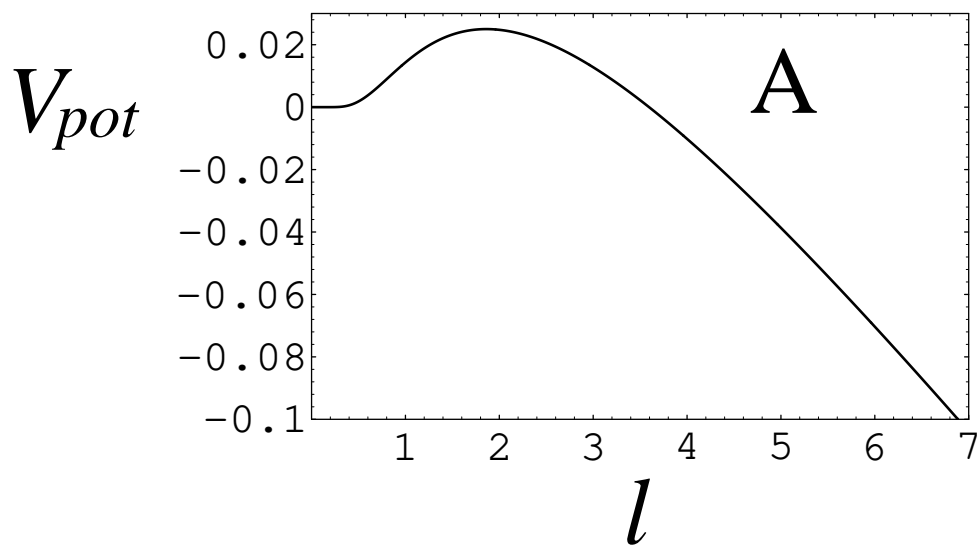
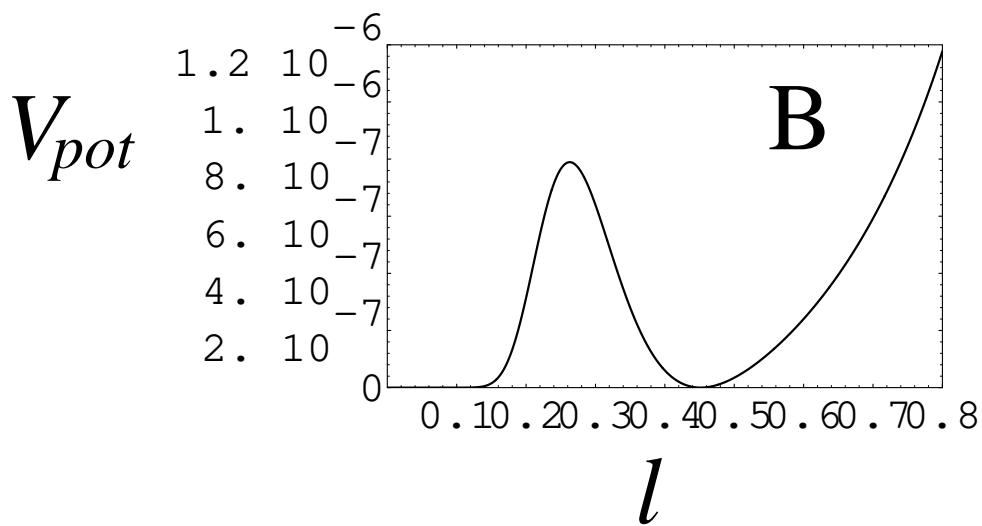
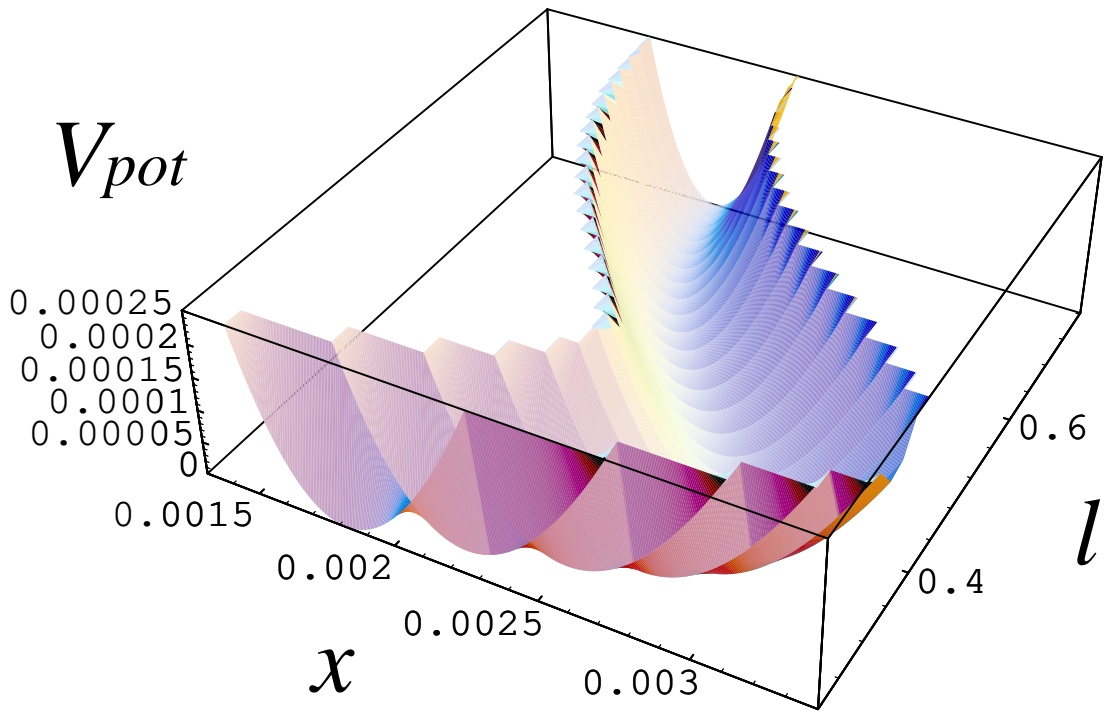
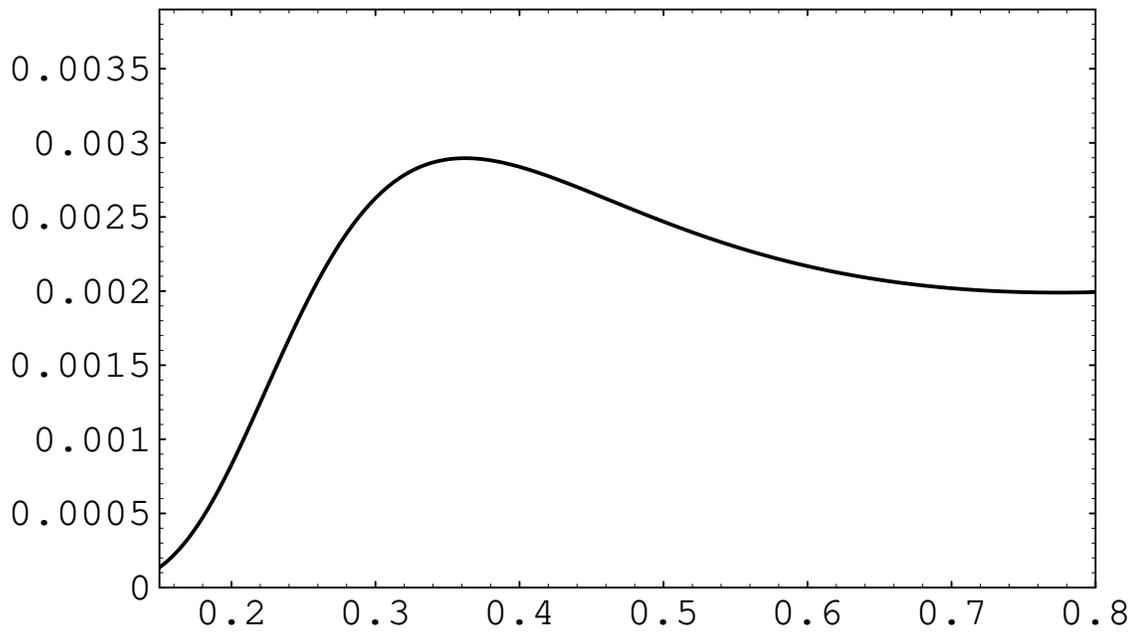
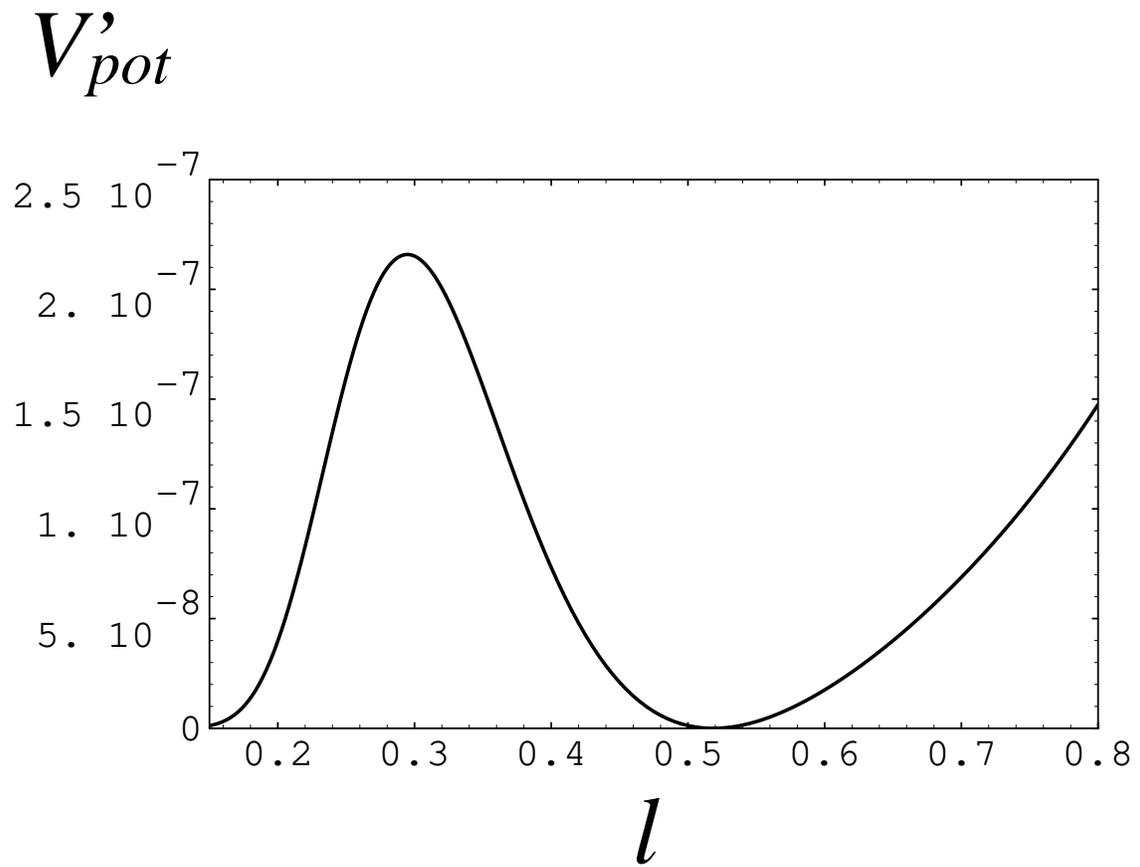
\end{document}